\documentclass[10pt,a4paper,twoside]{article}
\usepackage{epsfig}
\usepackage{baltlat6}
\usepackage{array}
\usepackage{here}
\begin{document}
\ \
\vspace{0.5mm}
\setcounter{page}{339}
\vspace{8mm}

\titlehead{Baltic Astronomy, vol.\,20, 339--354, 2011}

\titleb{A NEW DYNAMICAL PARAMETER FOR THE STUDY OF\\
STICKY ORBITS IN A 3D GALACTIC MODEL}

\begin{authorl}
\authorb{Euaggelos E. Zotos}{}
\end{authorl}

\moveright-3.2mm
\vbox{
\begin{addressl}
\addressb{}{Department of Physics,
Section of Astrophysics, Astronomy and Mechanics,\\
Aristotle University of Thessaloniki,
541 24, Thessaloniki, Greece;\\
evzotos@astro.auth.gr}
\end{addressl}  }

\submitb{Received: 2011 August 16; revised September 28; accepted:
October 10}

\begin{summary} A 3D dynamical model is used to study the motion in
the central parts of an elliptical galaxy, hosting a massive and dense
nucleus. Our aim is to investigate the regular or chaotic character of
the motion, with emphasis in the different chaotic components, as well
as the sticky regions of the dynamical system. In order to define the
character of the motion in the 2D system, we use the classical method
of the Poincar\'{e} $x-p_x$ phase plane, the Lyapunov Characteristic
Exponent (LCE) and the dynamical parameter -- the $S(c)$ spectrum. Then
the results obtained from the 2D system are used to investigate the
properties of the 3D system. For this, we introduce and use a new
dynamical parameter, the $S(k)$ spectrum, which proves to be a very
reliable and fast method to detect the islandic motion and the
evolution of the sticky orbits in the 3D system. Numerical experiments
conducted by the new $S(k)$ spectrum suggest that the different
chaotic components in the 3D system do not interact for time intervals
much larger than the age of the galaxy. The results indicate that the
different sticky regions do not lead to a unified chaotic sea. Thus,
the behavior of the 3D sticky orbits differs from that observed in the
2D system. Furthermore, the 3D motion near the center of a triaxial
elliptical galaxy seems to be very complicated, displaying several
families of resonant orbits, different chaotic components and sticky
regions, while only a small fraction of orbits is regular. The
comparison with earlier results reveals the importance of the
conception of the new dynamical spectrum. \end{summary}


\resthead{A new dynamical parameter for the study of 3D sticky orbits}
{Euaggelos E. Zotos}

\sectionb{1}{INTRODUCTION}

Now it is evident that elliptical galaxies hosting massive and dense
nuclei do exist (see Barth et al. 1999, 2001; Ho et al. 1995, 1997,
2000; Kaneda at al. 2005; Lauer at al. 2005; Maoz et al. 2005; Nagar et
al. 2005; Shields et al. 2000).  Therefore it is important to
investigate the dynamical behavior of stars in the central parts of
triaxial elliptical galaxies.  For the study of a real elliptical galaxy
with the particular overall density distribution, we choose a simple
dynamical model based on the well-known logarithmic potential (see
Binney \& Tremaine 2008).  We expand this potential in a Taylor series
near the origin $(x=0)$ with the terms up to the second degree in the
variables.  This expansion is valid only within certain distances from
the galaxy center.  In our case, the calculation of orbits has been done
up to the distances $R=\sqrt{x^2+y^2+z^2} \leq 1$ from the center,
hinting that only up to these distances the harmonic potential derived
from the logarithmic expansion is valid (see the Appendix for more
details about the Taylor expansion).

In the present study we investigate the motion of a test particle (star)
in a
3D gravitational galactic potential consisting of two parts:  the first
part is the potential of a 3D anisotropic harmonic oscillator (for
details see Kandrup \& Sideris 2002 and Kandrup \& Siopis 2003), derived
from the Taylor expansion of the logarithmic potential, and the second
part is a Plummer potential of a spherically symmetric nucleus.  Our 3D
dynamical model is
\begin{equation}
V(x,y,z) = \frac{\omega ^2}{2}\left(x^2 + a y^2 + b z^2 \right)
- \frac{M_n}{\sqrt{x^2 + y^2 + z^2 + c_n^2}} \ .
\end{equation}

The Plummer sphere which describes the potential of the spherical
nucleus, has been applied several times in the past to study the effect
of the introduction of a central mass component in a galaxy (see Hasan
\& Norman 1990; Hasan et al. 1993).  Here $a$ and $b$ are parameters,
$M_n$ and $c_n$ are the mass and the scale length of the nucleus, while
the parameter $\omega$ is used for the consistency of the galactic
units.  We use a system of galactic units, where the unit of mass is
$2.325 \times 10^7$ M$_\odot$, the unit of length is 1 kpc and the unit
of time is $0.997748 \times 10^8$ yr.  The velocity unit is 10 km/s,
while $G$ is equal to unity.  In the above units we use the values:
$\omega$ =10 km\,s$^{-1}$\,kpc$^{-1}$, $a=4$, $b=1.25$, $M_n= 50$, while
$c_n$ is treated as a parameter.  The test particle (star) which moves
under the influence of the gravitational potential (1) is considered to
have the mass $m=1$.

There are two basic reasons justifying the choice of the potential (1).
First, this simple potential can reasonably describe the character of
orbits in the central parts of a triaxial elliptical galaxy.  The second
and most important reason is that the above potential displays important
characteristics of the motion, such as a large variety of resonant
orbits, different chaotic components and remarkable sticky regions.
Therefore, it seems challenging to investigate the nature of the above
characteristics, in order to draw some useful conclusions regarding the
behavior of orbits in the central region of a triaxial elliptical
galaxy, hosting a dense and massive nucleus.

The Hamiltonian to the potential (1) writes
\begin{equation}
H = \frac{1}{2} \left( p_x^2 + p_y^2 + p_z^2 \right) + V(x,y,z) = h \ ,
\end{equation}
 where $p_x, p_y$ and $p_z$ are the momenta per unit mass conjugate to
 $x, y$ and $z$ respectively, while $h$ is the numerical value of the
 Hamiltonian. The outcomes of this research are based on the numerical
integration of the equations of motion
\begin{eqnarray}
\dot{x}&=&p_x, \ \ \ \dot{y}=p_y, \ \ \ \dot{z}=p_z \ , \nonumber \\
\dot{p_x}&=&-\frac{\partial \ V(x,y,z)}{\partial x} \ , \nonumber\\
\dot{p_y}&=&-\frac{\partial \ V(x,y,z)}{\partial y} \ , \nonumber\\
\dot{p_z}&=&-\frac{\partial \ V(x,y,z)}{\partial z} \ ,
\end{eqnarray}
where the dot indicates derivative with respect to the time.  We use a
Bulirsh -- St\"{o}er integration routine in Fortran 95, with double
precision in all subroutines.  The accuracy of the outcomes was checked
by constancy of the energy integral (2), which was conserved up to the
fifteenth significant figure.

In order to distinguish between the regular or chaotic motion in
different chaotic components, we use the Lyapunov Characteristic
Exponent (LCE), see Lichtenberg \& Lieberman (1992).  Moreover, in
order to be able to visualize and study the sticky and islandic motion
in the 2D system, we use a dynamical parameter -- the $S(c)$ spectrum.
Dynamical spectra have been frequently used over the last years for
obtaining fast and reliable results regarding the regular or chaotic
nature of orbits.  Recently, new definitions of the dynamical spectra
have been introduced in order to identify the regular or chaotic nature
of orbits in the Hamiltonian systems of two (2D) and three (3D) degrees
of freedom (see Zotos 2011a,b).

Here we must remind to the reader that this spectrum is the distribution
function of the parameter $c$
\begin{equation}
S(c)=\frac{\Delta N(c)}{N \Delta c} \ ,
\end{equation}
 where $\Delta N(c)$ is the number of the parameters $c$ in the
 interval $\left(c, c+\Delta c \right)$ after $N$ iterations. The
parameter $c$ is defined as
\begin{equation}
c_i=\frac{x_i-p_{xi}}{p_{yi}} \ ,
\end{equation}
where $(x_i,p_{xi},p_{yi})$ are the successive values of the
$(x,p_x,p_y)$ elements of a 2D orbit on the Poincar\'{e} $x-p_x$, $y=0$,
$p_y>0$ phase plane.  More details regarding the shapes and behavior of
dynamical spectra can be found in Caranicolas \& Papadopoulos
(2007) and Caranicolas \& Zotos (2010).

The layout of this article is as follows:  in Section 2 we study the
behavior of orbits in the 2D system.  In the same Section a systematic
presentation of sticky regions is made, and the evolution of sticky
orbits is investigated using the $S(c)$ spectrum.  Furthermore, we study
the degree of chaos in the different chaotic components of the model
applying the values of LCE.  We try to find out the role of the scale
length $c_n$ of the nucleus in the evolution of different chaotic
components.  In Section 3 we investigate a 3D model, using a new
dynamical indicator, the $S(k)$ spectrum.  Here we are interested to
find out if the different chaotic components in the 2D system merge in
order to form a 3D unified chaotic manifold.  Another interesting
question is to determine the initial conditions in the $H\left(x, p_x,
z\right) = h, p_y>0, \left(y=p_z=0\right)$ phase space, giving rise to
regular or chaotic orbits.  Finally, in Section 4 we present a
discussion and the conclusions of this research.

\sectionb{2}{THE STRUCTURE OF THE 2D DYNAMICAL SYSTEM}

Let us now proceed to study the character of orbits in a 2D model.
The corresponding 2D Hamiltonian is
\begin{equation}
H_2 = \frac{1}{2} \left( p_x^2+p_y^2 \right) + V(x,y)=h_2 \ ,
\end{equation}
where $h_2$ is the numerical value of Hamiltonian (6).  Figure 1 shows
the $x-p_x$ phase plane for this Hamiltonian when $h_2=1.5$.  The
values of the parameters are:  $\omega = 10$, $a=4$, $b=1.25$, $M_n=50$
and $c_n=0.25$.  As one can see, this is a very interesting and
complicated phase plane, with considerable sticky regions, sets of
islands of invariant curves produced by a number of secondary
resonances and several
different chaotic components.  In order to help the reader, we will
describe in detail each of the above cases.


\begin{figure}[!tH]
\centering
\resizebox{0.80\hsize}{!}{\rotatebox{0}{\includegraphics*{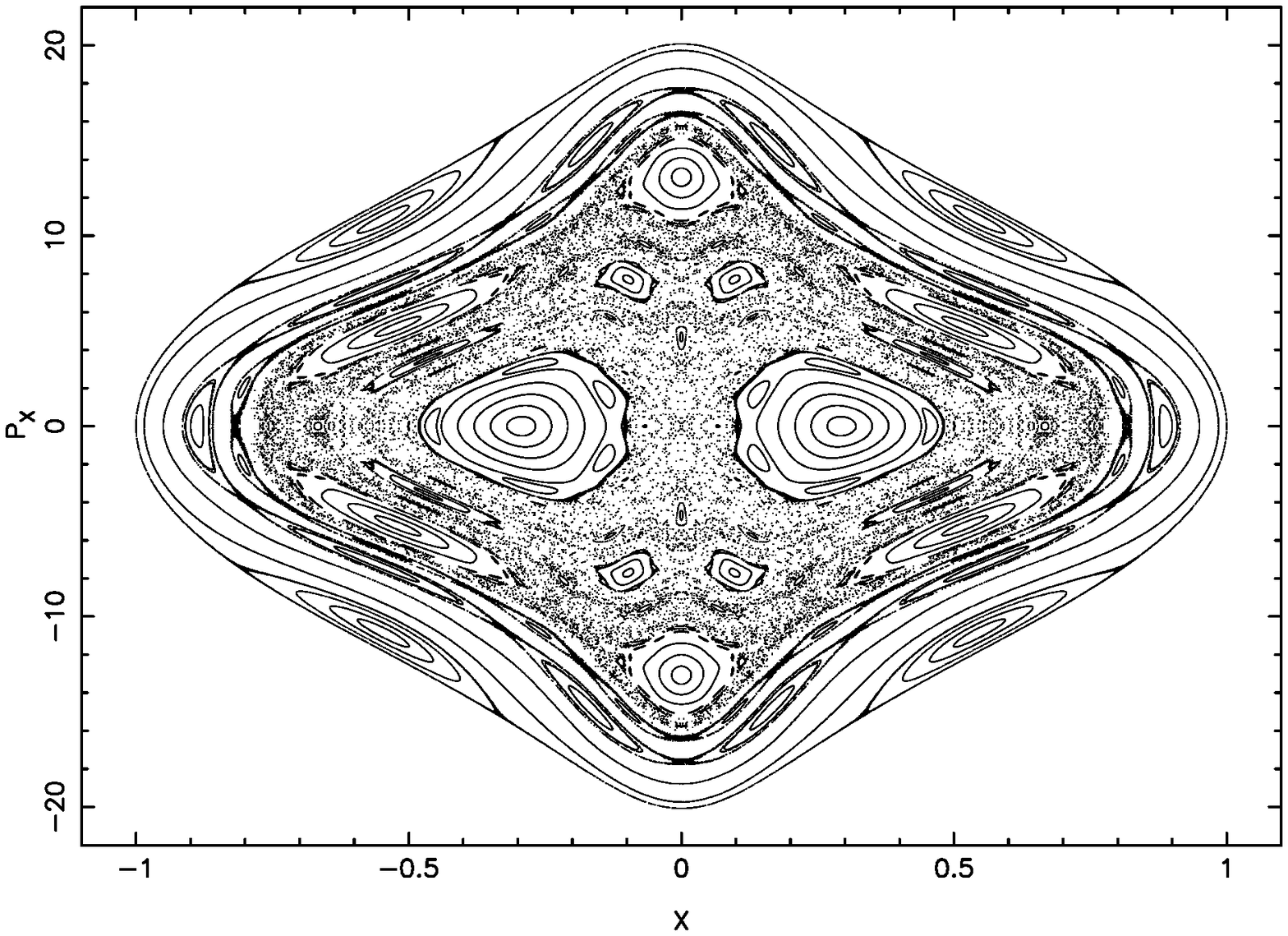}}}
  \captionb{1}{The $x-p_x$ phase plane, when $h_2=1.5$. Details are
given in the text.}
\end{figure}

Figures 2\,(a)--(d) show the three sticky regions observed in the
$x-p_x$
phase plane of the dynamical system.  Figure 2a shows the sticky region
I, consisting of two large sticky islands.  The initial conditions are:
$x_0= \pm 0.1$, $p_{x0}=0$ and the integration time is 2000 time units.
Figure 2b shows the sticky region II consisting of eight islands.  The
initial conditions are:  $x_0=0.06$, $p_{x0}=6.77$ and the integration
time is 900 time units.  Note that the four sticky islands are produced
by one sticky orbit, while the other four are produced by the twin
symmetric sticky orbit traversed in the opposite direction.  Figure 2c
shows the sticky region III composed of six sticky islands produced by
the two twin orbits.  The initial conditions are:  $x_0=0$,
$p_{x0}=10.5$ and the integration time is 750 time units.  Each sticky
orbit produces three sticky islands.  Figure 2d shows all three sticky
regions which are embedded in the chaotic sea.  This chaotic sea is the
place, where the test particle (star) ends after each sticky period.
The
integration time of each region shown in Figures 2\,(a--d) is equal to
each sticky period.


\begin{figure*}[!tH]
\begin{center}
\resizebox{\hsize}{!}{\rotatebox{0}{\includegraphics*{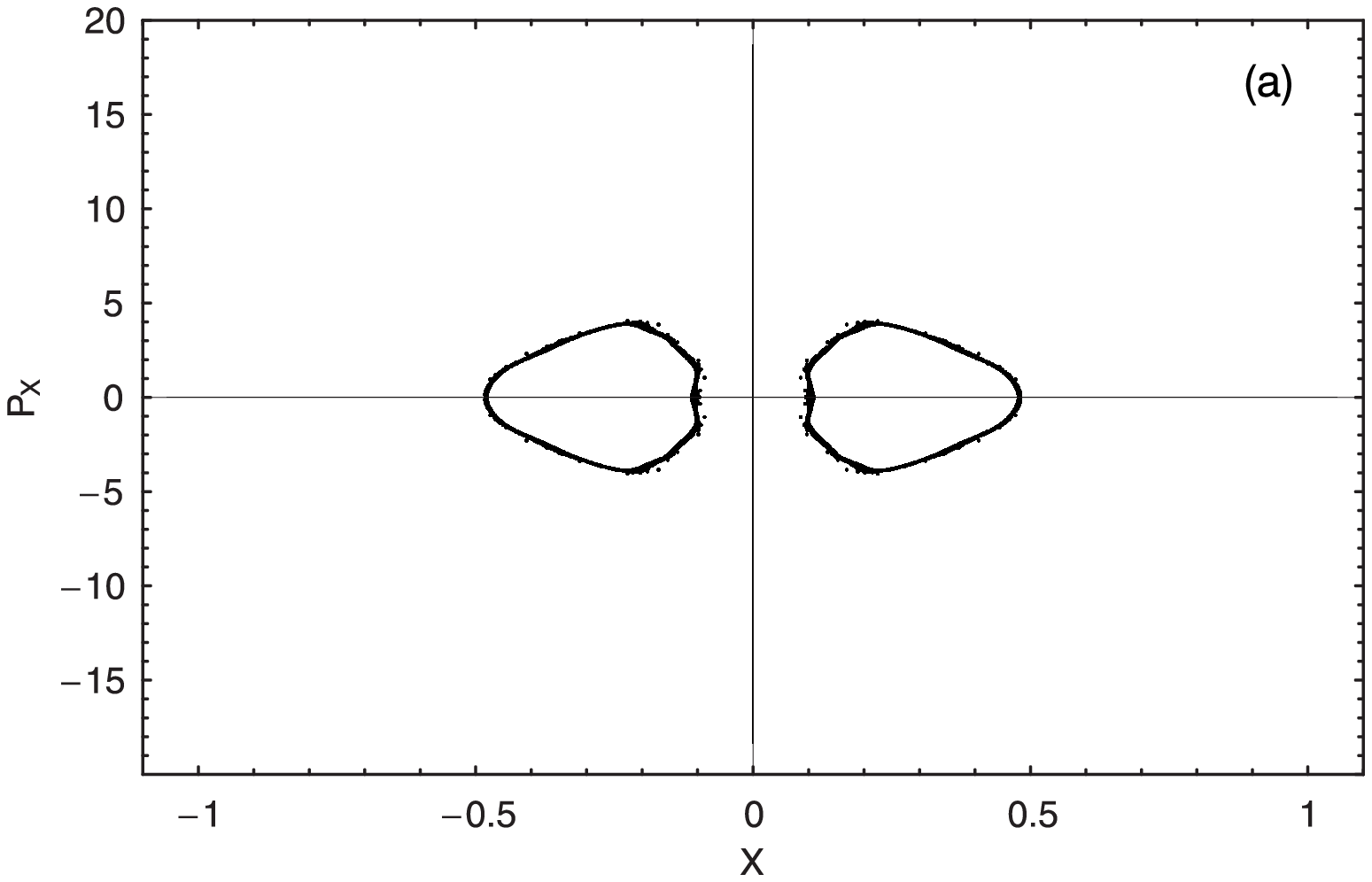}}\hspace{2cm}
                      \rotatebox{0}{\includegraphics*{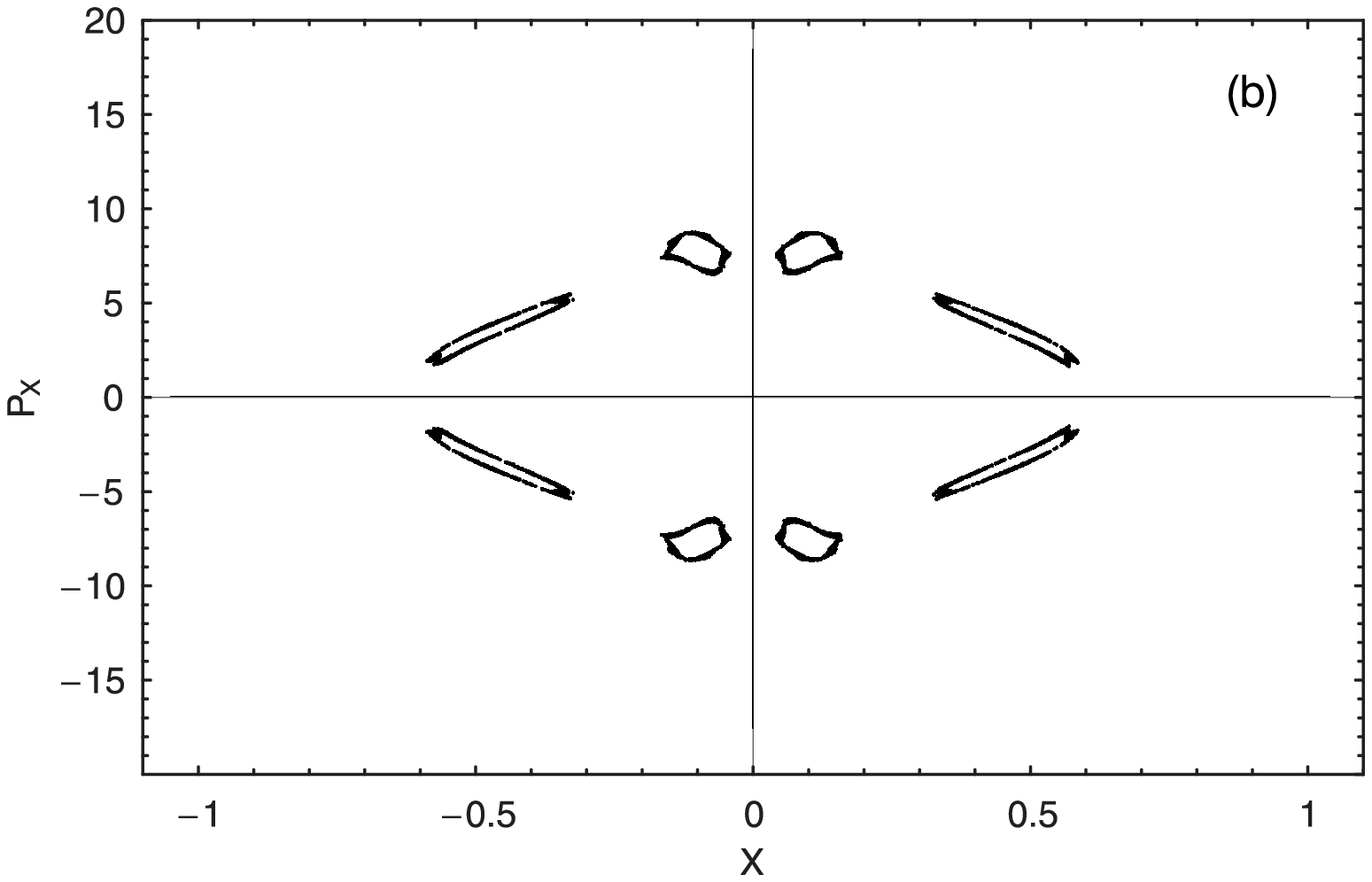}}}
\resizebox{\hsize}{!}{\rotatebox{0}{\includegraphics*{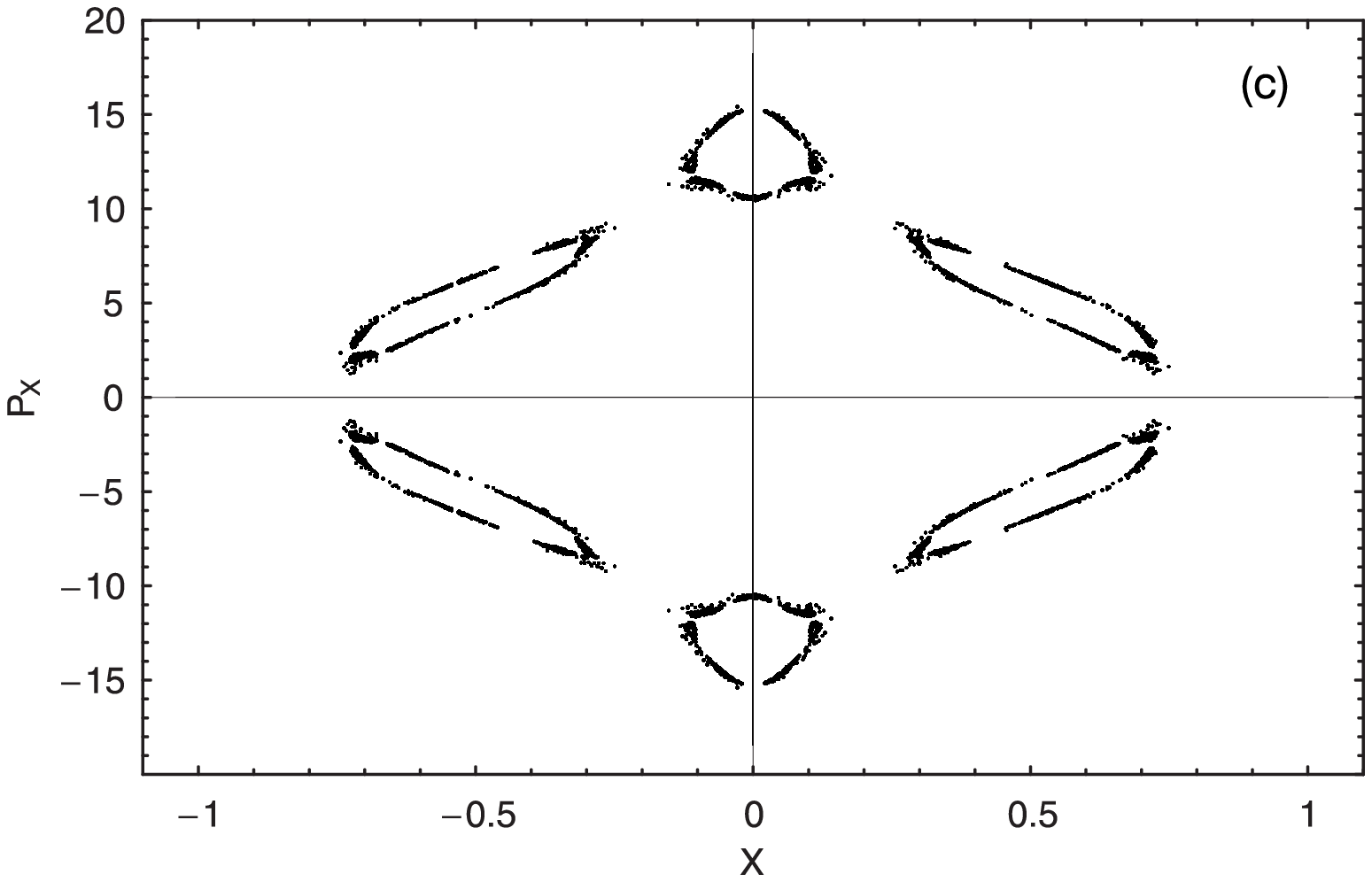}}\hspace{2cm}
                      \rotatebox{0}{\includegraphics*{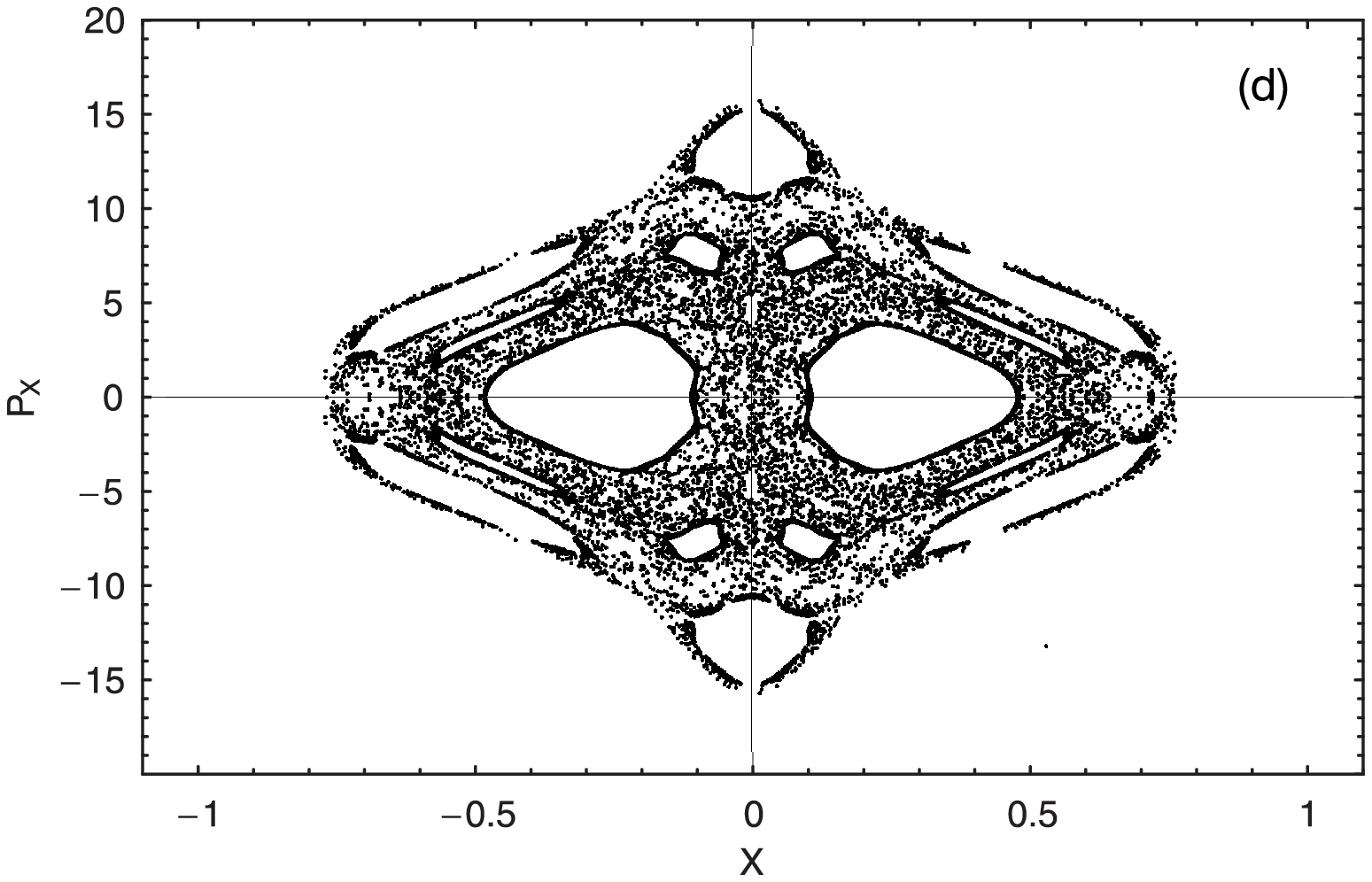}}}
\end{center}
\vskip 0.02cm
\captionb{2}{Panels (a)--(c): the three sticky regions and
panel (d): the sticky regions embedded in the chaotic sea.}
\vspace{5mm}
\end{figure*}


\begin{figure*}[!tH]
\begin{center}
\resizebox{\hsize}{!}{\rotatebox{0}{\includegraphics*{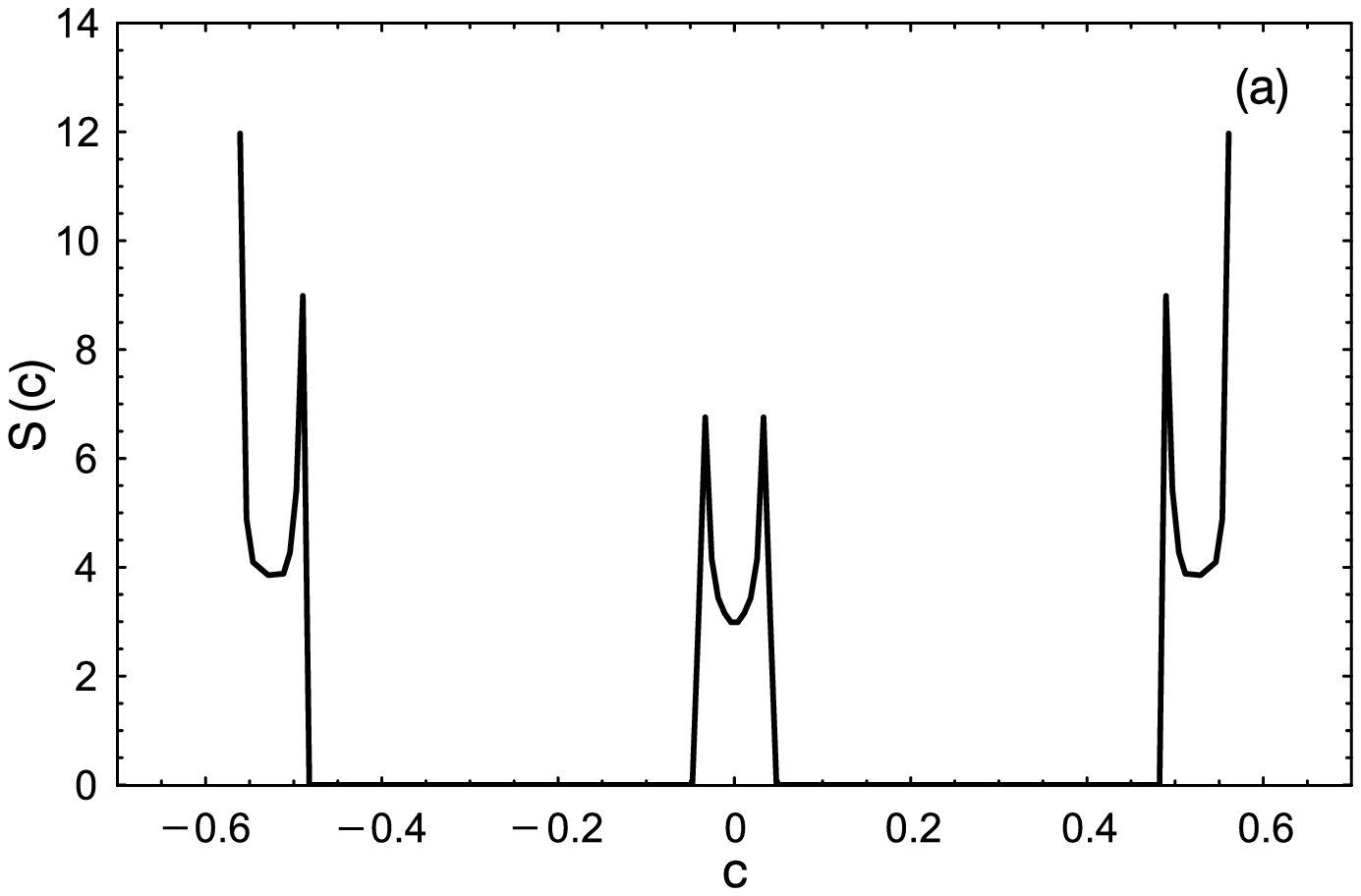}}\hspace{2cm}
                      \rotatebox{0}{\includegraphics*{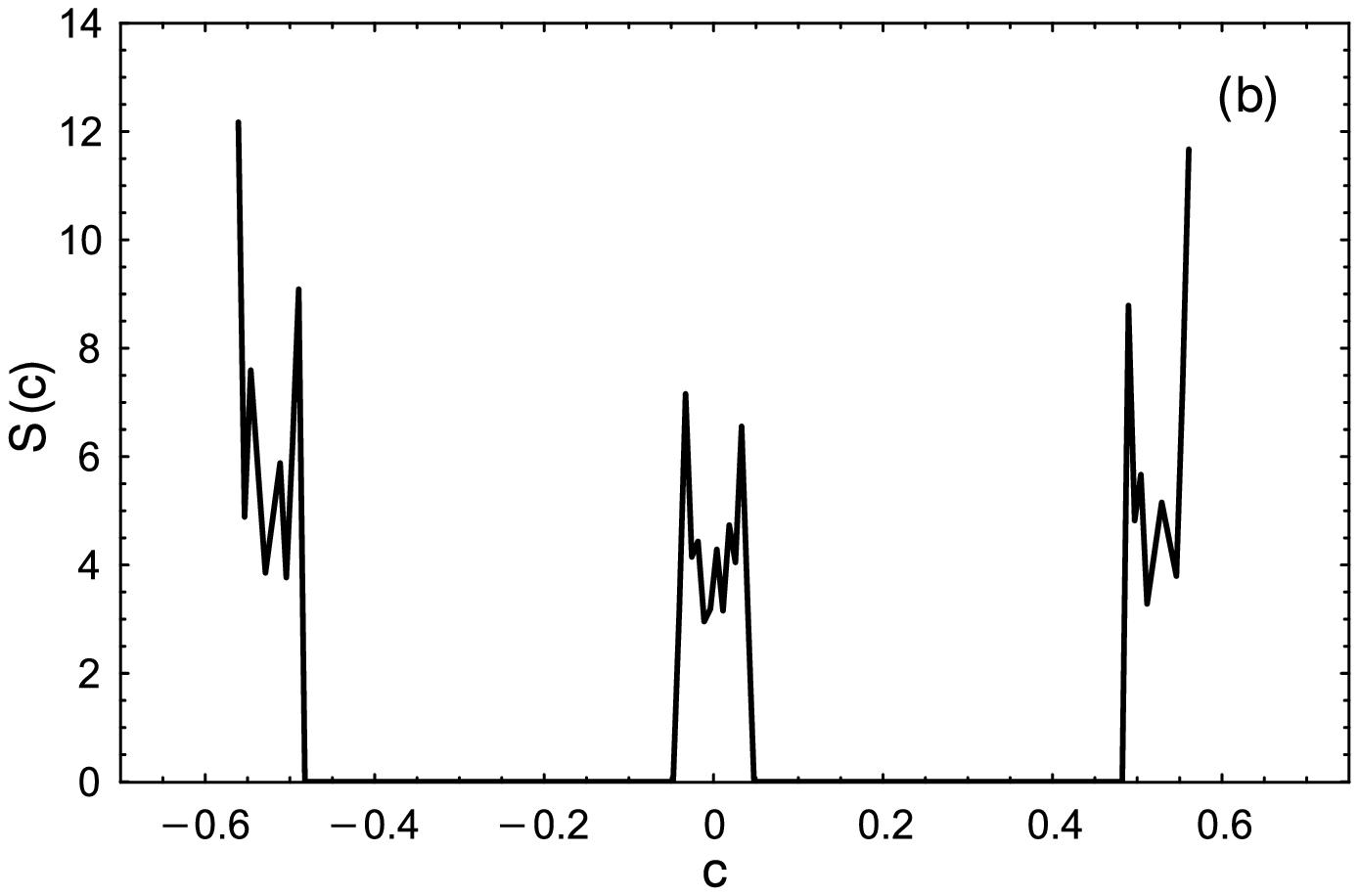}}}
\resizebox{\hsize}{!}{\rotatebox{0}{\includegraphics*{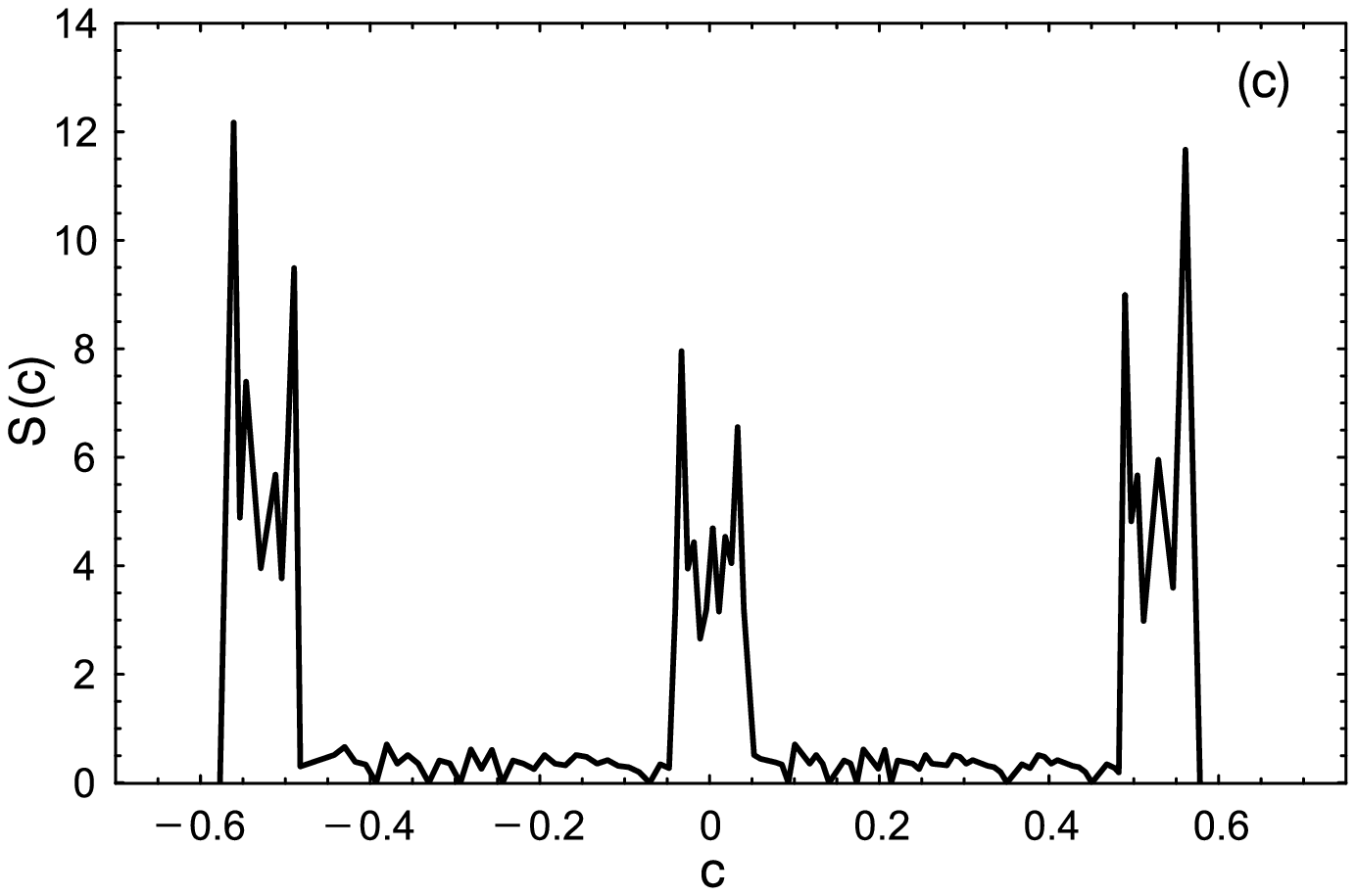}}\hspace{2cm}
                      \rotatebox{0}{\includegraphics*{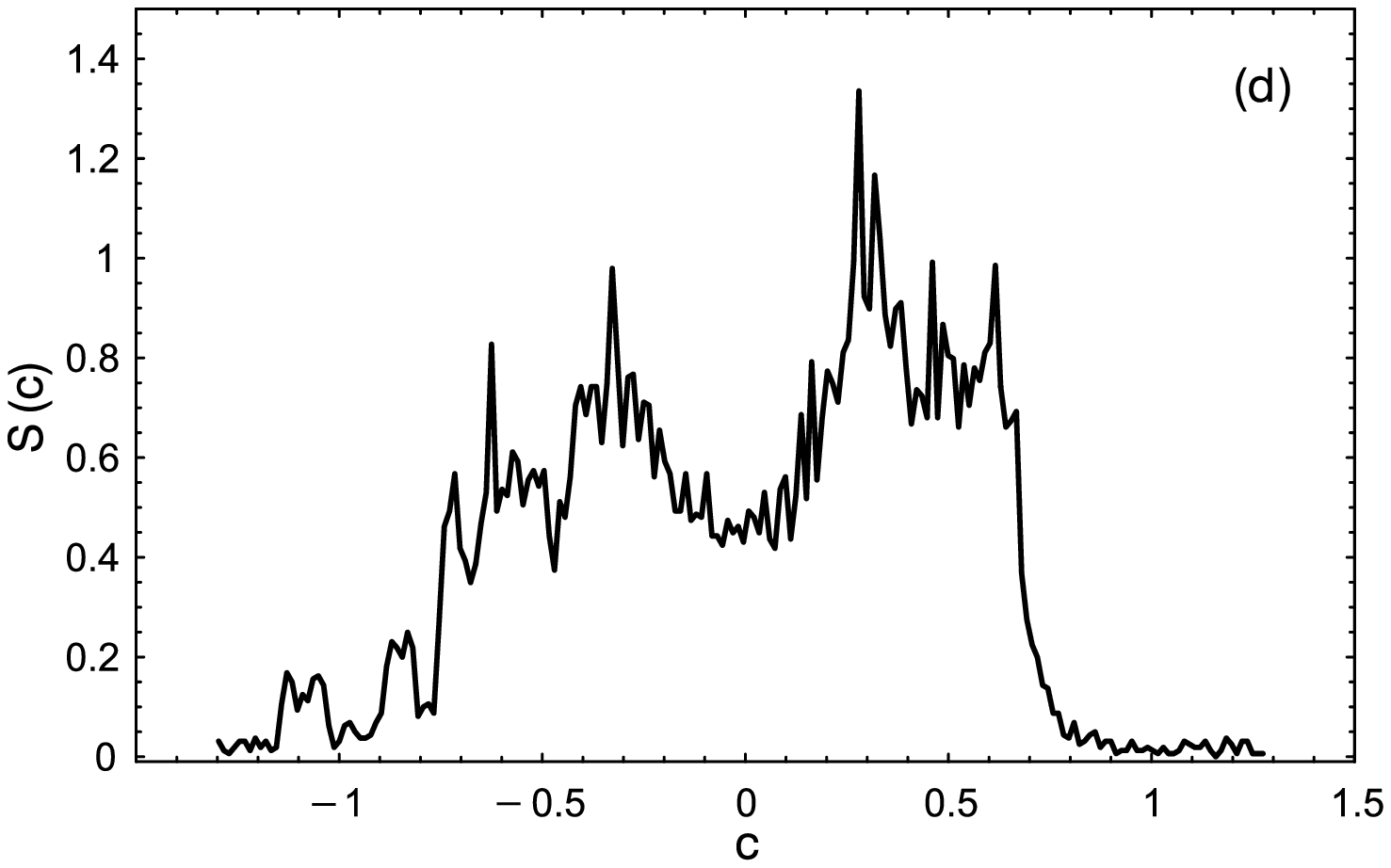}}}
\end{center}
\vskip 0.02cm
\captionb{3}{Panel (a): the $S(c)$ spectrum of a
nearby quasi-periodic orbit and panels (b)--(d): evolution of the $S(c)$
spectrum for a sticky orbit.}
\end{figure*}

The evolution of sticky orbits can be followed using the $S(c)$
spectrum.  The results are shown in Figures 3\,(a)--(d).  Figure 3a
shows the spectrum of an orbit producing three islands.  The initial
conditions are:  $x_0=0$, $y_0=0$, $p_{x0}=12.6$, while the value of
$p_{y0}$ is always found from the energy integral (6).  The values of
all other parameters are as in Figure 1. Here we observe three well
defined $U$-type spectra, indicating a regular motion.  Note that the
number of spectra is equal to the number of islands.  Figure 3b shows
the $S(c)$ spectrum of an orbit starting near the above regular orbit.
The initial conditions are:  $x_0=0$, $y_0=0$, $p_{x0}=10.5$.  Here we
observe three different spectra with a number of large and small peaks.
This indicates that we have a sticky orbit, and the sticky region is
composed of the three sticky islands.  The sticky period is about
$T=720$ time units.  Figure 3c shows the spectrum of the same orbit when
$T=780$.  Here, the three spectra have merged to produce a unified
spectrum.  This indicates that after the sticky period the test particle
(star) has moved to the chaotic sea.  Figure 3d shows the spectrum of
the orbit when $T=3000$.  Here we observe the spectrum of a chaotic
orbit with a large number of small and large asymmetric peaks.  Note
that the sticky period obtained from the $S(c)$ spectrum is very close
to the integration time of Figure 2c.  The $S(c)$ spectrum shows that in
all sticky regions of the dynamical system the sticky period is between
700--900 time units and finally the test particle (star) ends in the
chaotic sea surrounding all the three sticky regions.


\begin{figure*}[!tH]
\begin{center}
\resizebox{\hsize}{!}{\rotatebox{0}{\includegraphics*{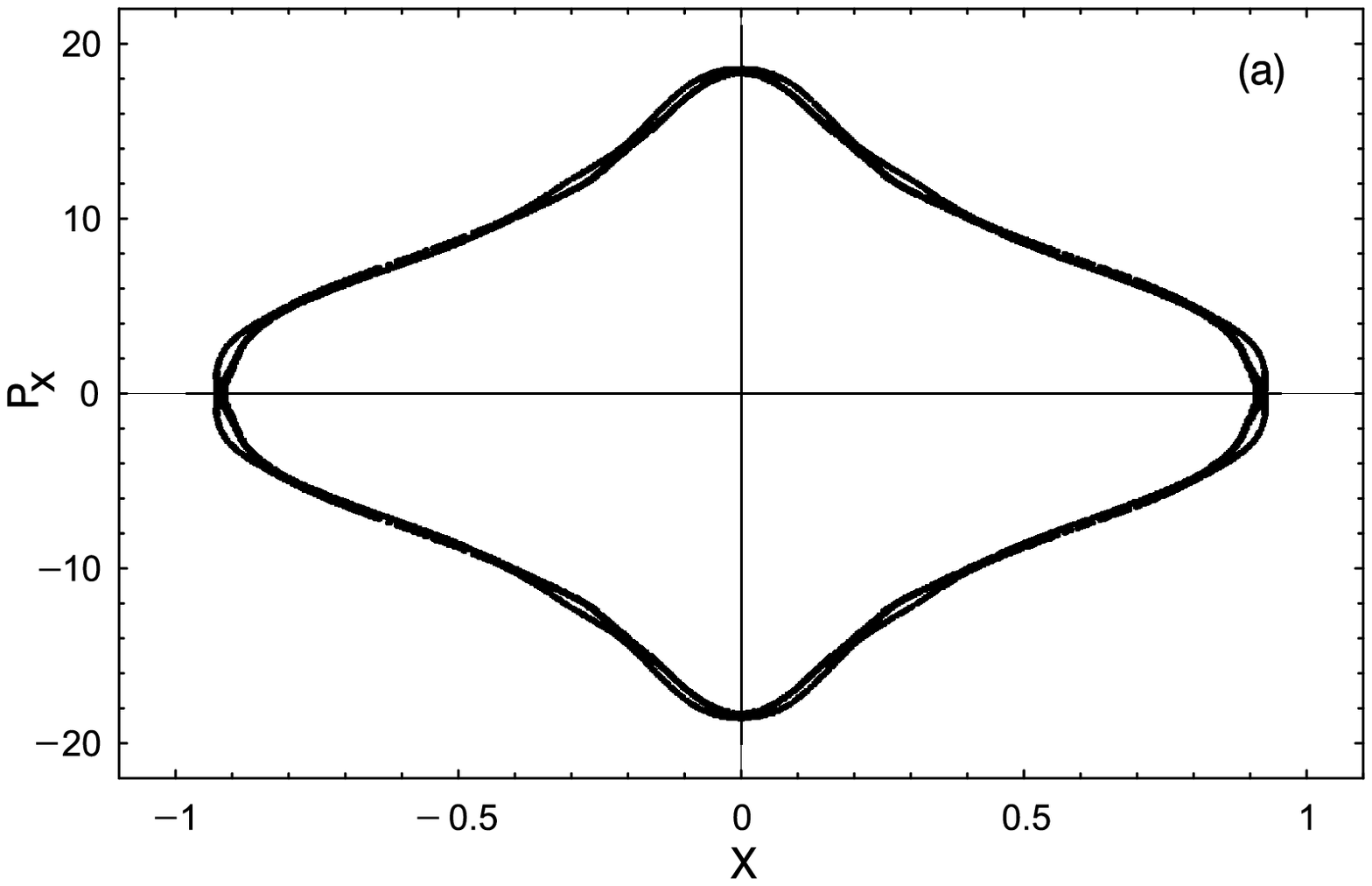}}\hspace{2cm}
                      \rotatebox{0}{\includegraphics*{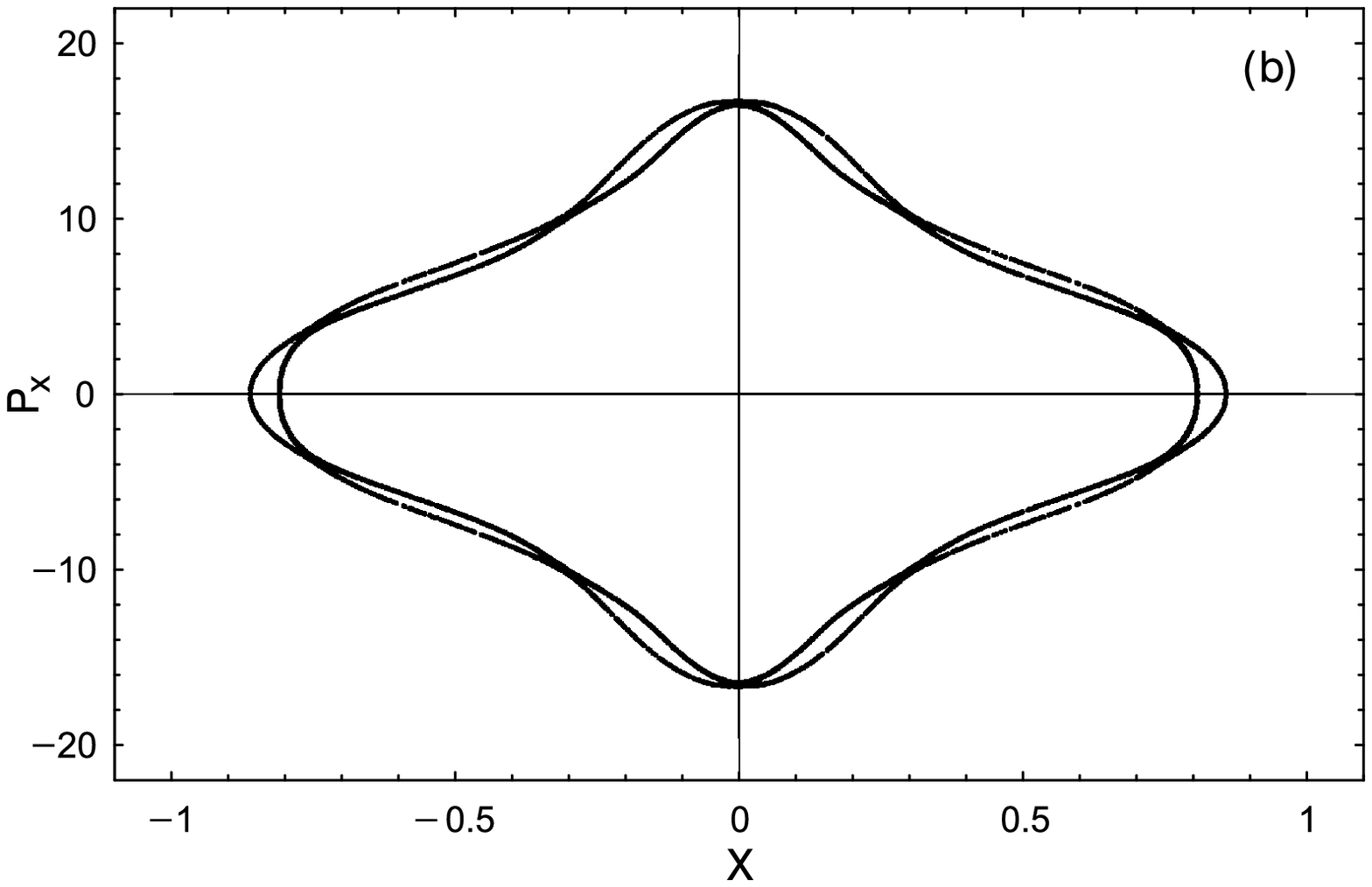}}}
\resizebox{\hsize}{!}{\rotatebox{0}{\includegraphics*{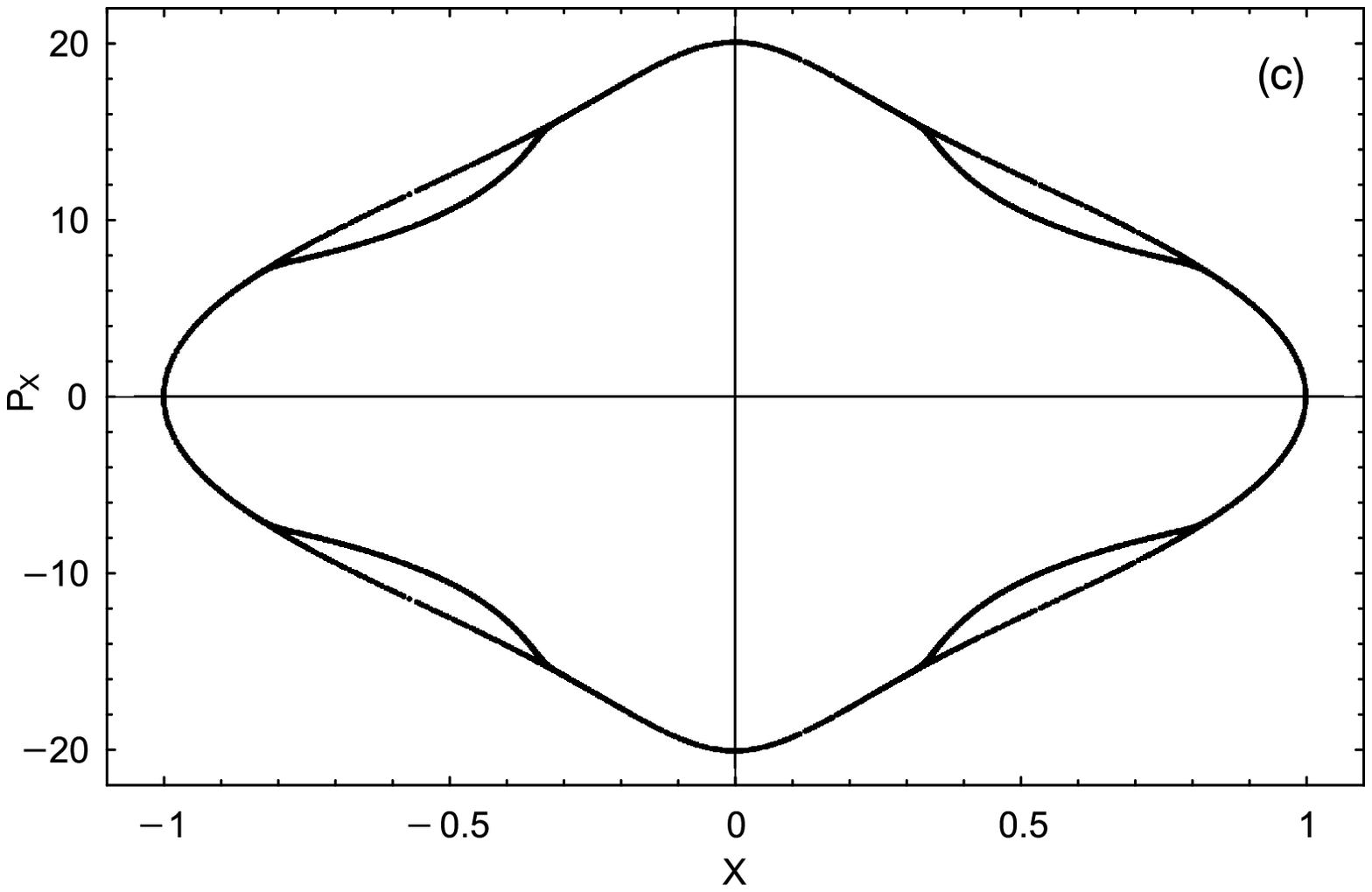}}\hspace{2cm}
                      \rotatebox{0}{\includegraphics*{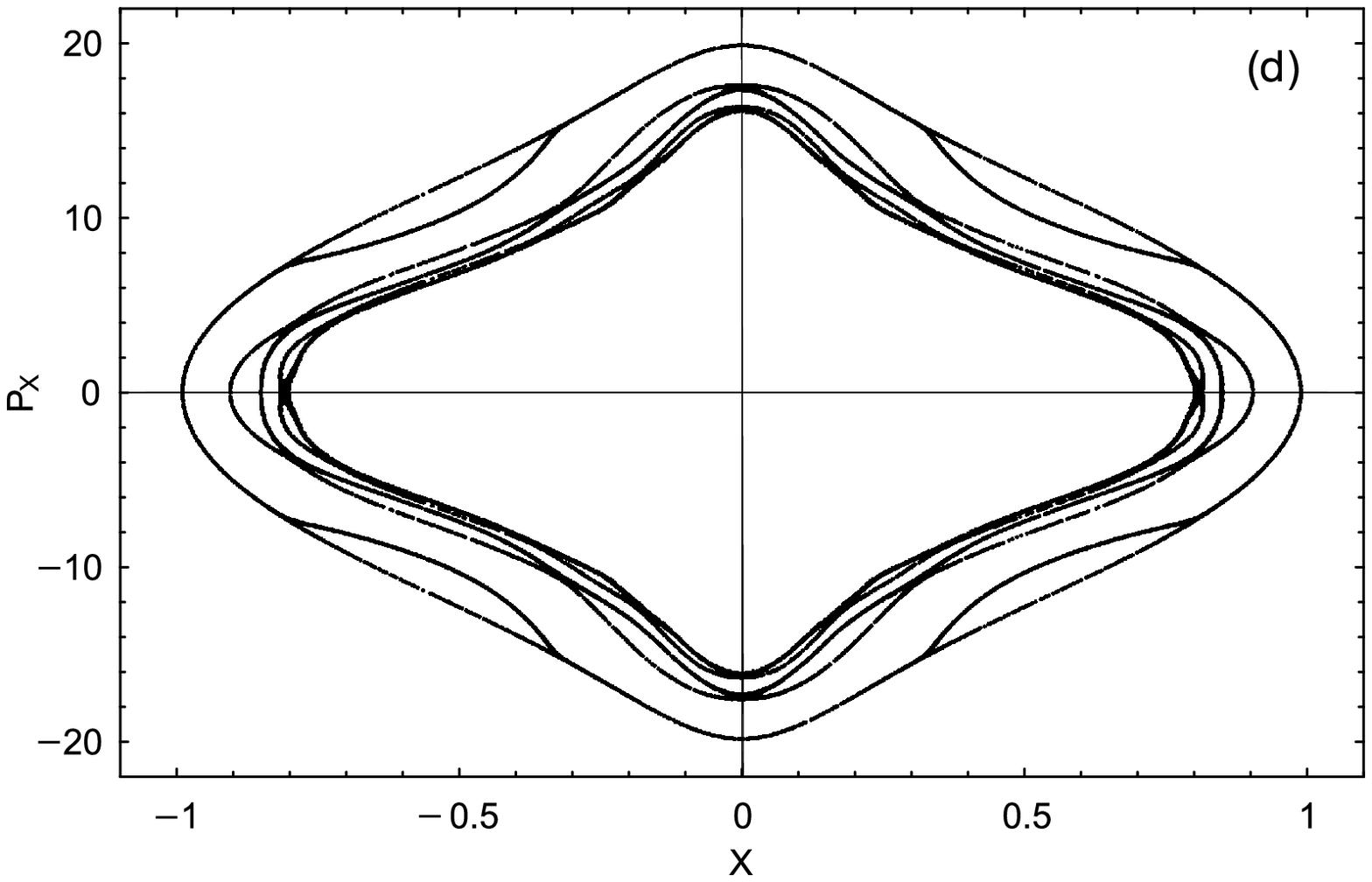}}}
\end{center}
\vskip 0.02cm
\captionb{4}{Panels (a)--(c): the three chaotic components and
panel (d): all the chaotic components together.}
\end{figure*}


\begin{figure}[!tH]
\begin{center}
\resizebox{0.80\hsize}{!}{\rotatebox{0}{\includegraphics*{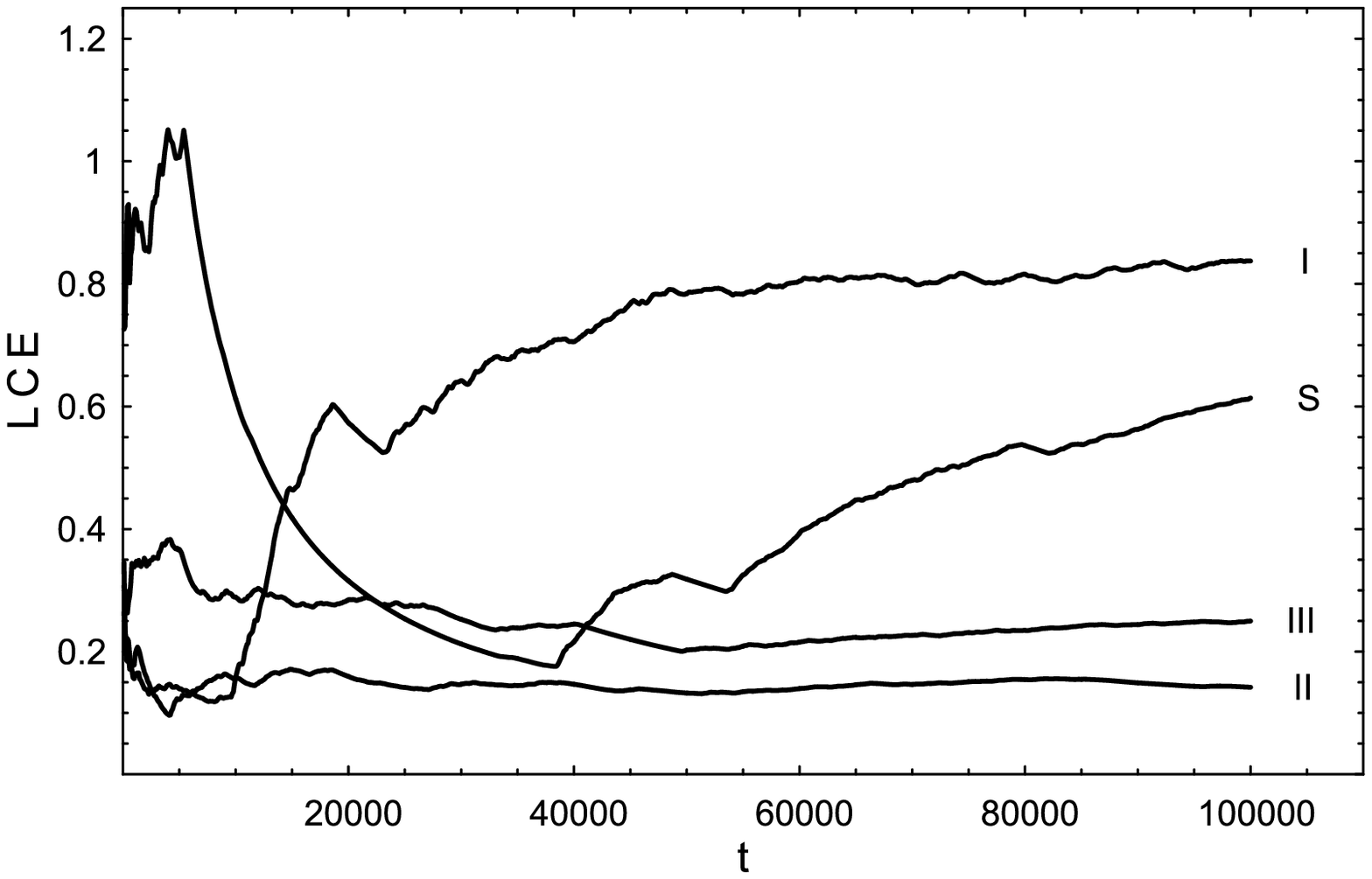}}}
\end{center}
\captionb{5}{The LCEs for the three different chaotic components I, II,
III and the chaotic sea S.}
\end{figure}


\begin{figure*}[!tH]
\begin{center}
\resizebox{\hsize}{!}{\rotatebox{0}{\includegraphics*{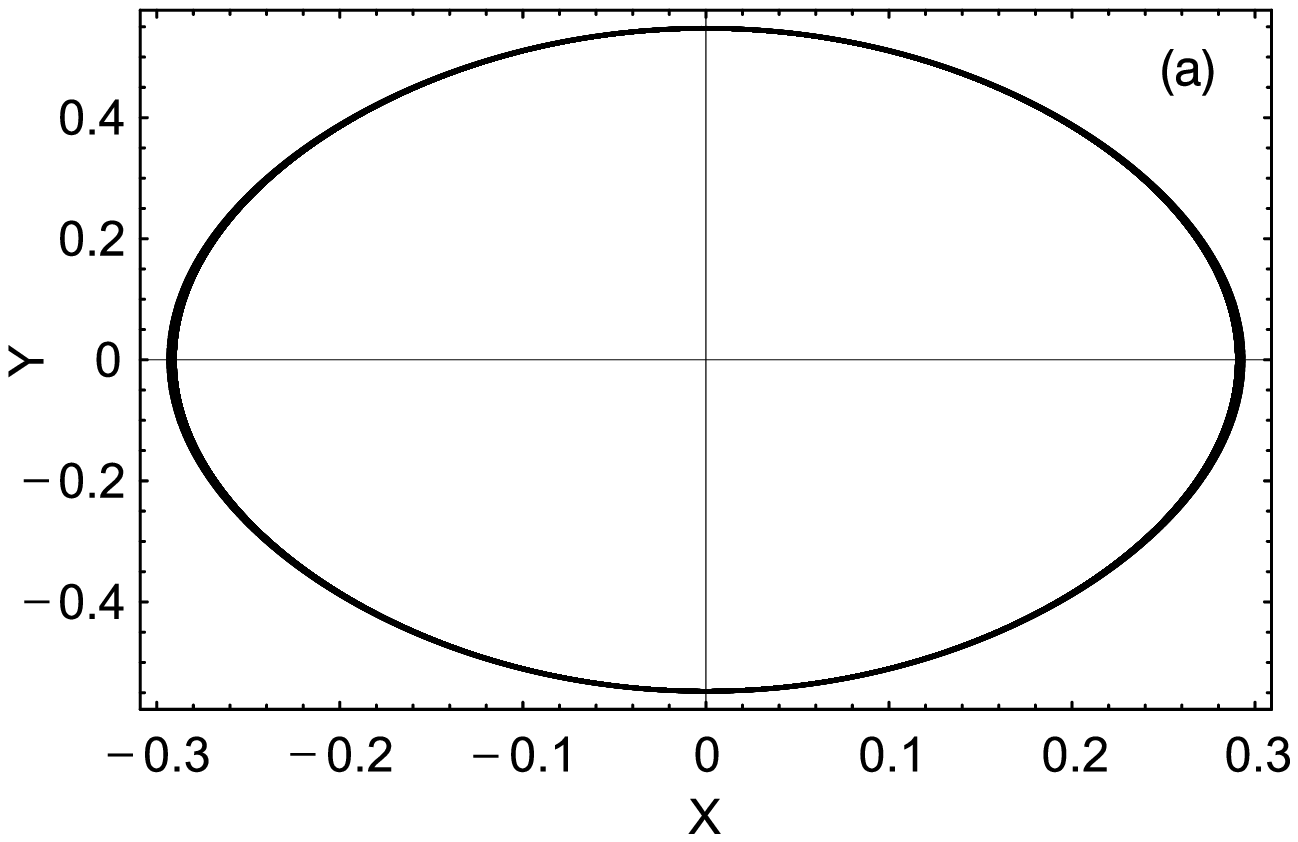}}\hspace{2cm}
                      \rotatebox{0}{\includegraphics*{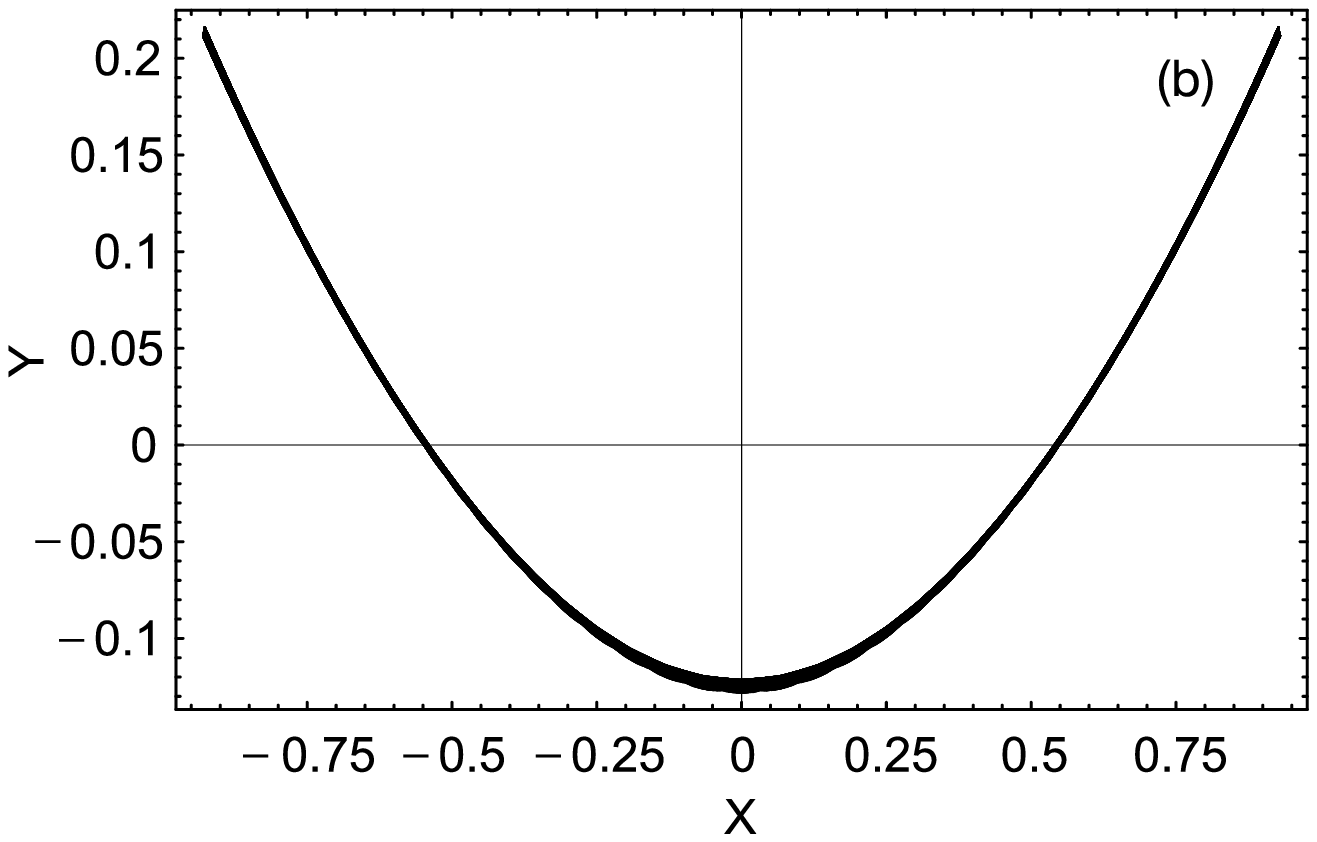}}}
\resizebox{\hsize}{!}{\rotatebox{0}{\includegraphics*{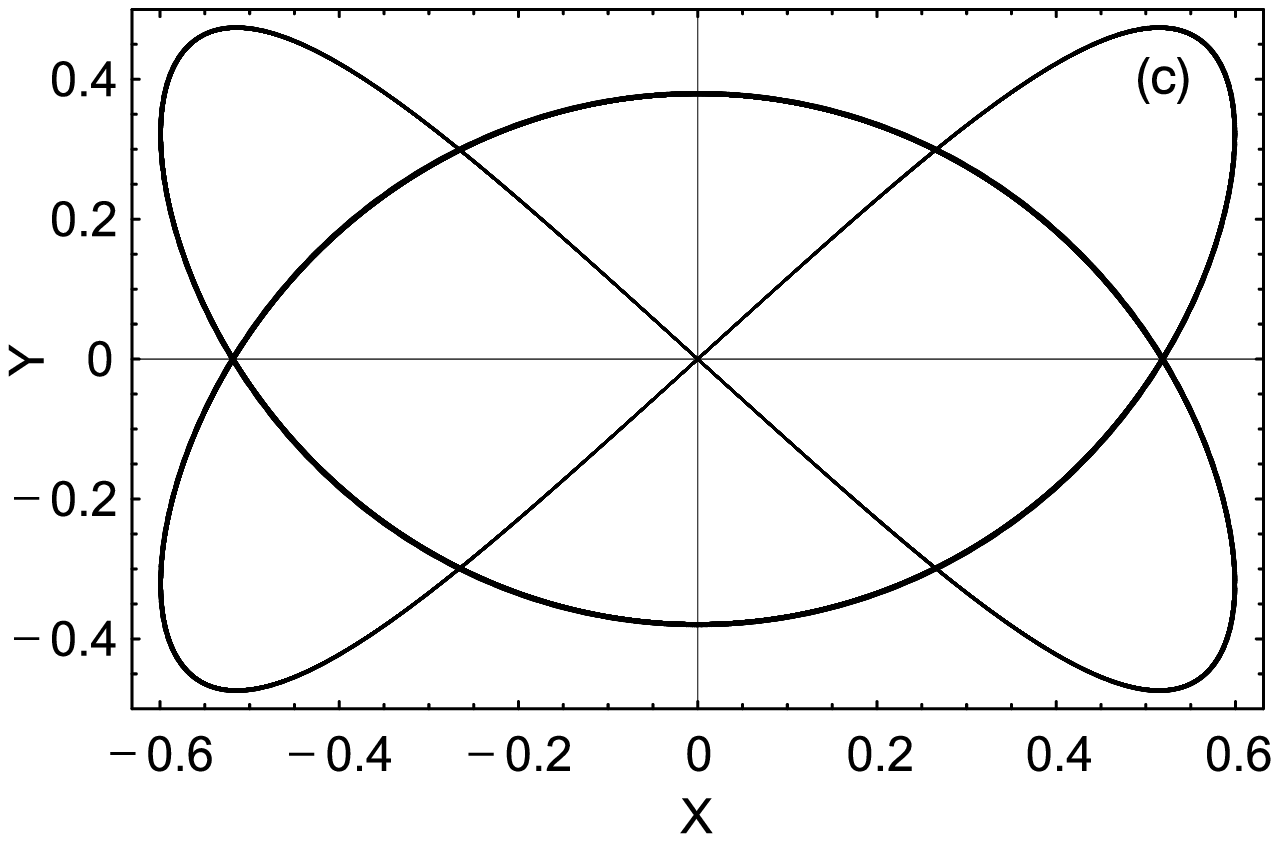}}\hspace{2cm}
                      \rotatebox{0}{\includegraphics*{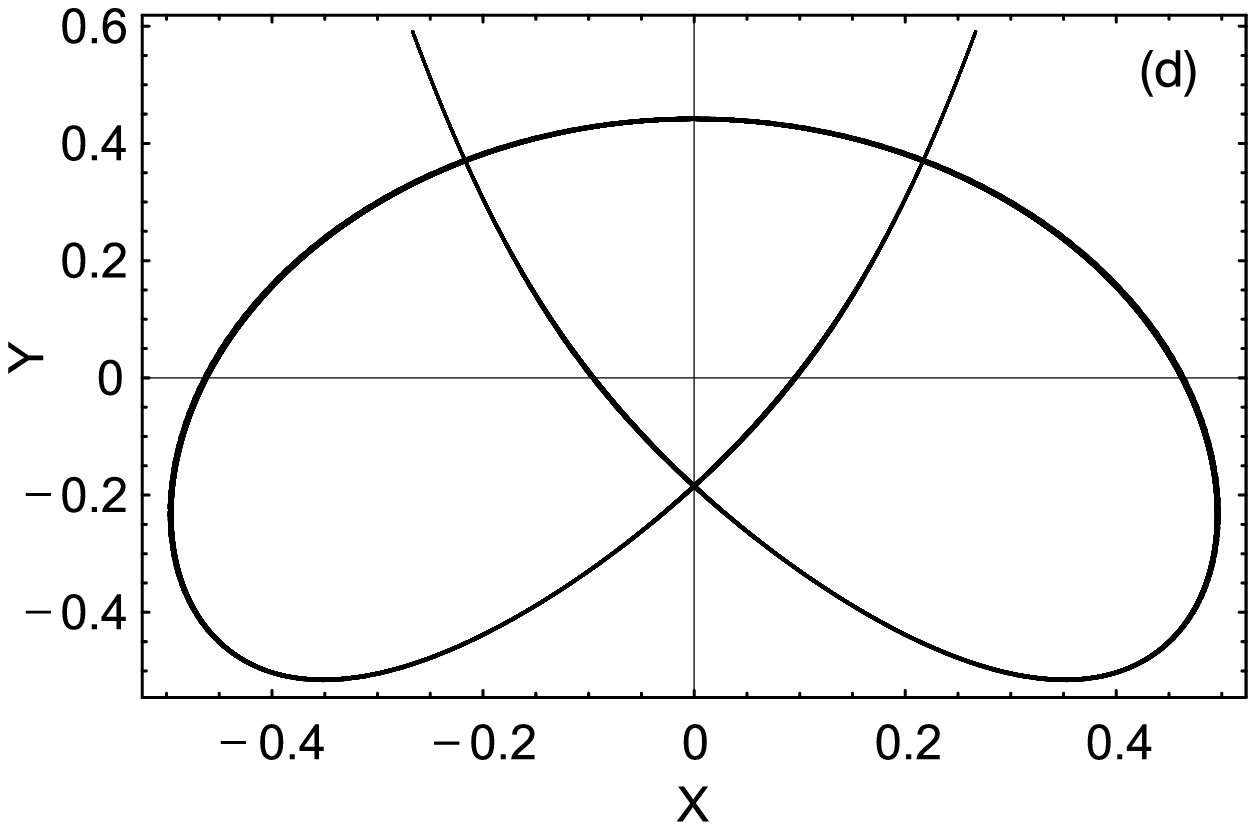}}}
\resizebox{\hsize}{!}{\rotatebox{0}{\includegraphics*{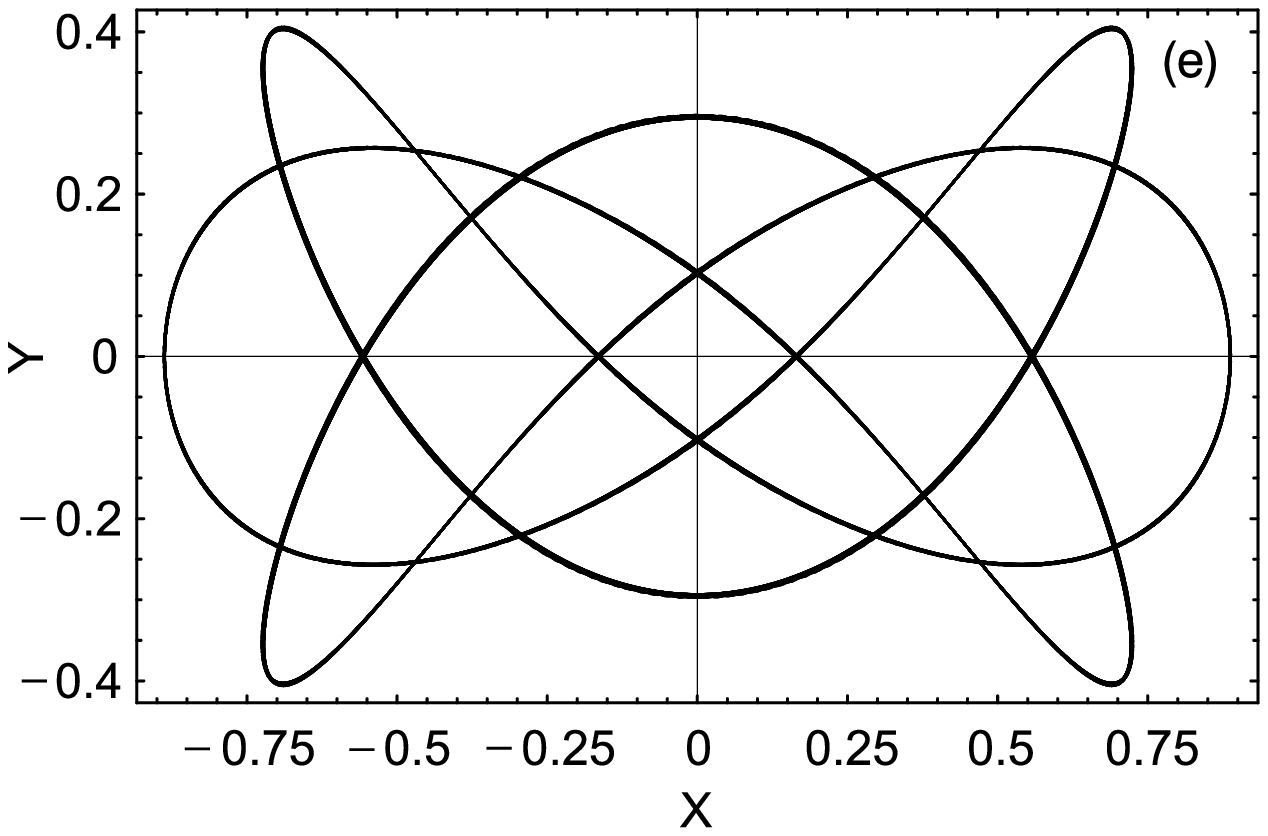}}\hspace{2cm}
                      \rotatebox{0}{\includegraphics*{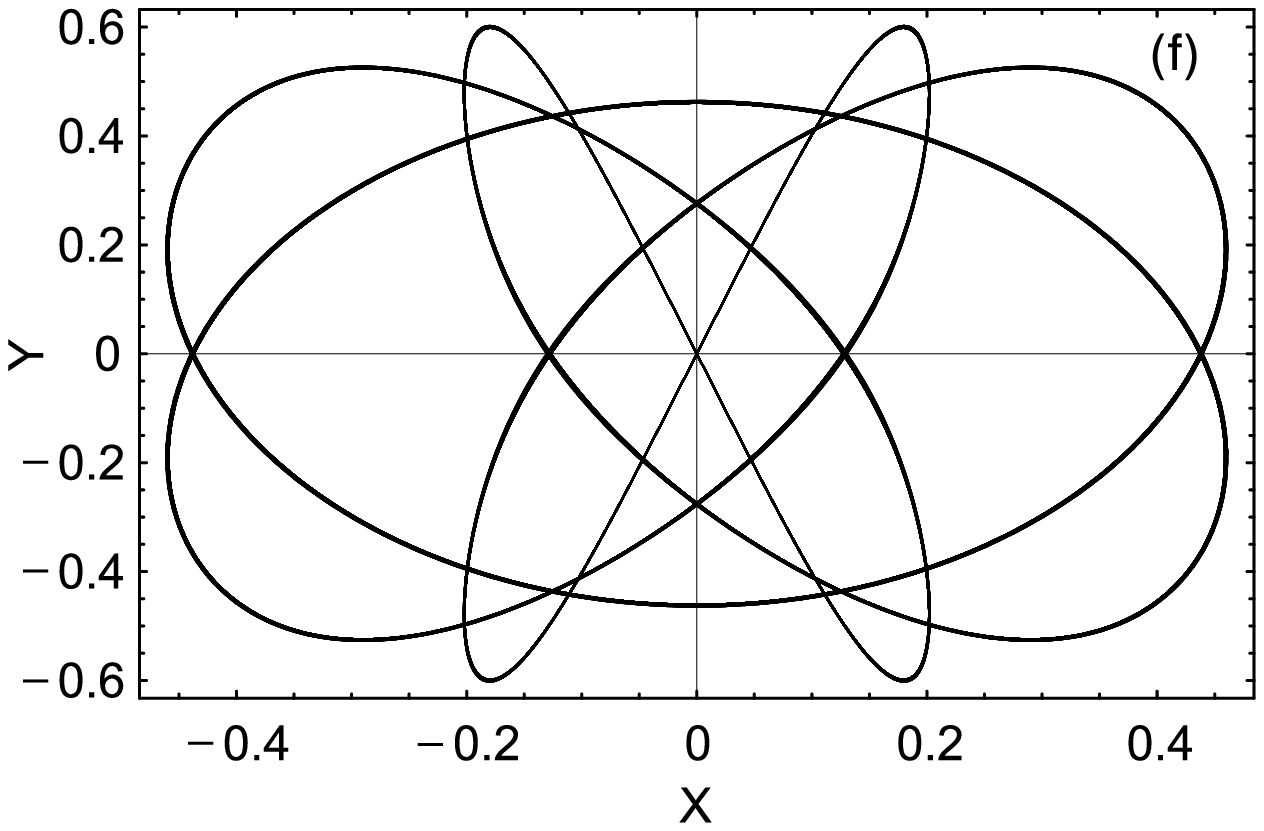}}}
\resizebox{\hsize}{!}{\rotatebox{0}{\includegraphics*{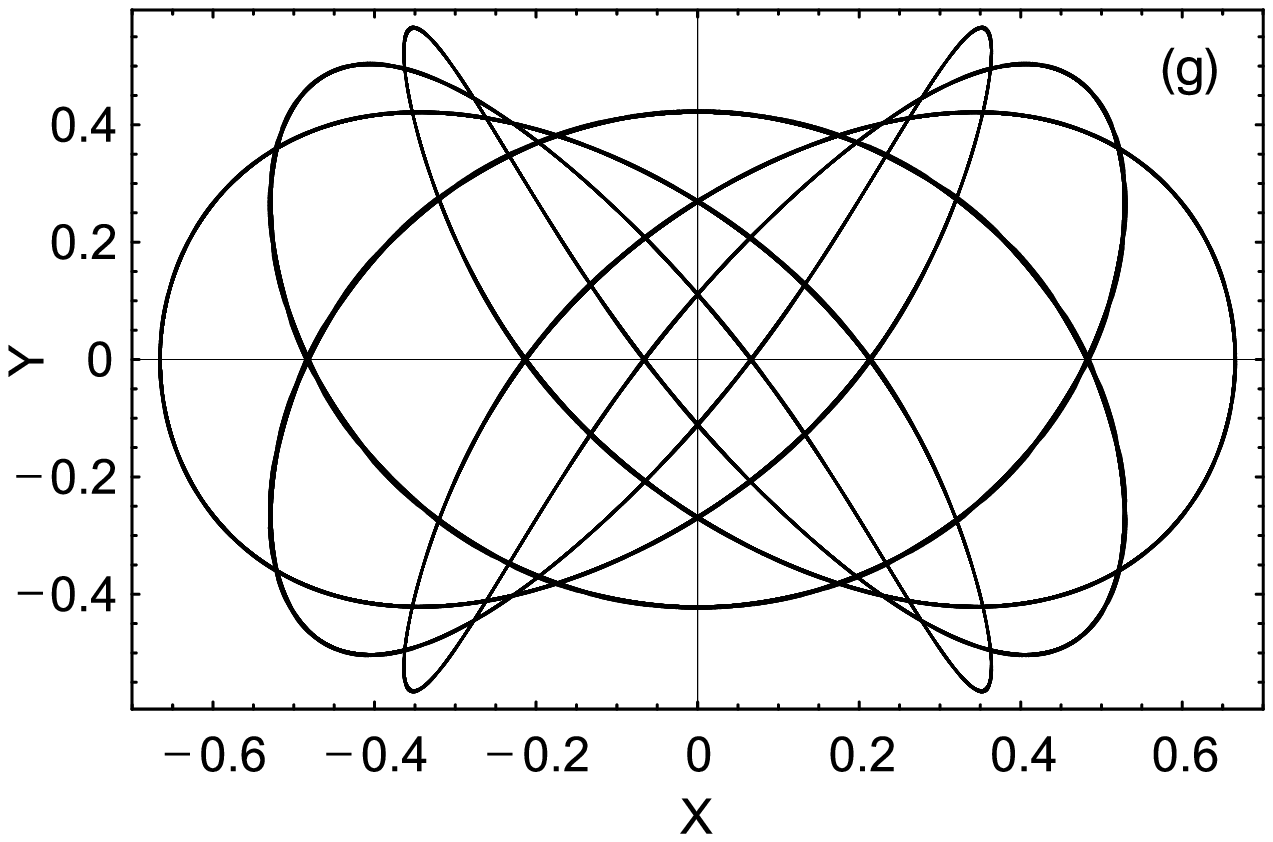}}\hspace{2cm}
                      \rotatebox{0}{\includegraphics*{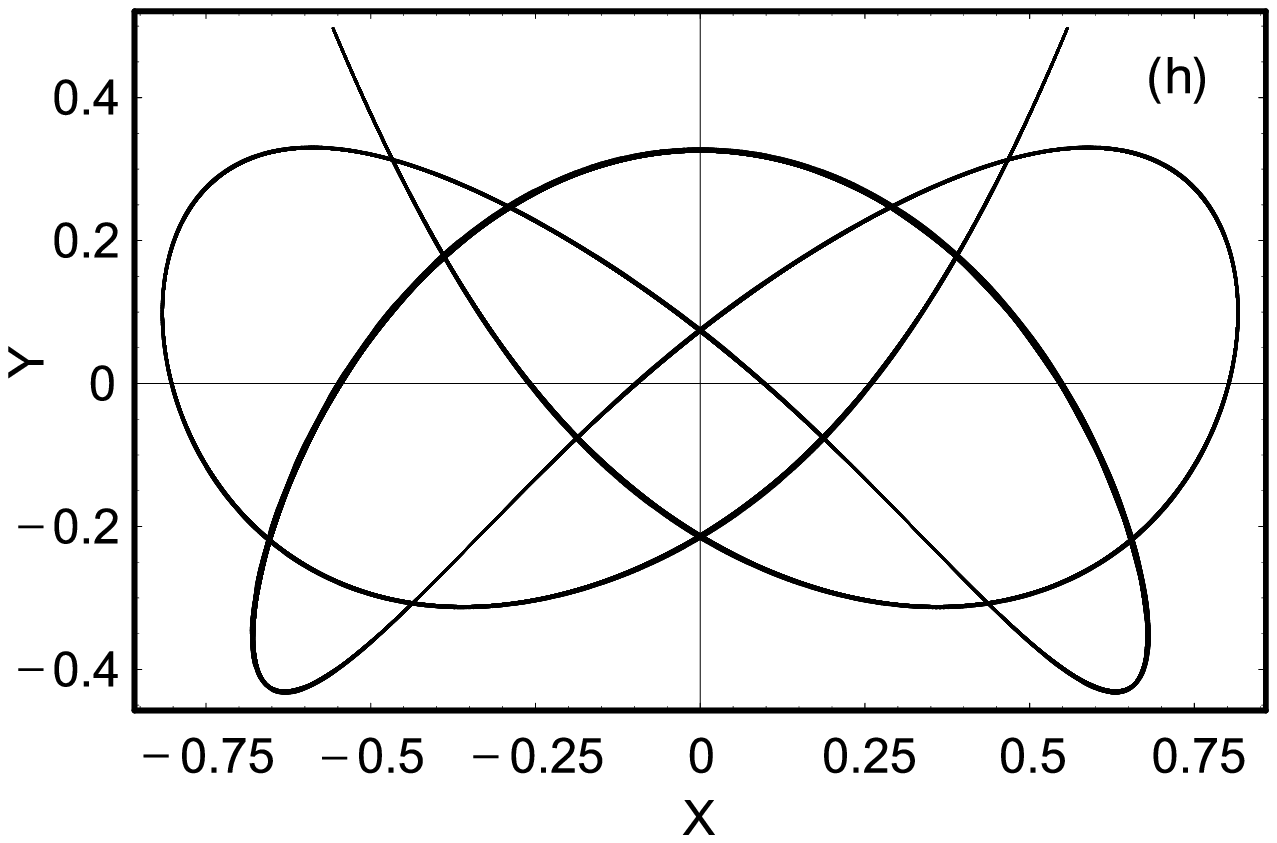}}}
\end{center}
\vskip 0.02cm
\captionb{6}{Panels (a)--(h): representative periodic orbits in the 2D
dynamical
system. The initial conditions are (a): $x_0=0.2943$, $p_{x0}=0$; (b):
$x_0=0.548$, $p_{x0}=10.8$; (c): $x_0=0$, $p_{x0}=13.07$; (d):
$x_0=0.097$, $p_{x0}=7.7$; (e): $x_0=0.889$, $p_{x0}=0$; (f): $x_0=0$,
$p_{x0}=4.675$; (g): $x_0=0.667$, $p_{x0}=0$ and (h): $x_0=0.803$,
$p_{x0}=2.03$.}
\end{figure*}

Figures 4\,(a)--(d) show the different chaotic components observed in
the
$x-p_x$ phase plane of the dynamical system.  Figure 4a shows the
chaotic
component I with the initial conditions $x_0=-0.82$ and $p_{x0}=0$.
Figure 4b shows the chaotic component II with the initial conditions
$x_0=-0.86$ and $p_{x0}=0$.  Figure 4c shows the chaotic component III
with the initial conditions $x_0=-1$ and $p_{x0}=0$.  All the three
chaotic components together are shown in Figure 4d.  The above three
chaotic components, together with the chaotic sea shown in Figure 2d,
are the four chaotic components of the 2D system.  The integration time
of the chaotic components, shown in Figures 4\,(a)--(d), is 2000 time
units.  As expected, each chaotic component has its own value of LCE
(see Saito \& Ichimura 1979).  A plot of LCEs for a time period of
$10^5$ time units is shown in Figure 5. I, II, III indicate the three
chaotic components, while S stands for the chaotic sea. The LCE for the
chaotic component I has the value 0.8372, for the chaotic component II
the value 0.1423 and for the chaotic component III the value 0.2512.
The LCE value for the chaotic sea is 0.6137.

Figures 6\,(a)--(h) show eight typical periodic orbits for the 2D
dynamical system.  The values of all other parameters are as in
Figure~1.  Figure 6a shows a periodic orbit characteristic to the 1:1
resonance, while the orbits shown in Figures 6b, 6c, 6d, 6e, 6f, 6g and
6h are periodic orbits characteristic to the 1:2, 2:3, 3:4, 3:5, 4:5,
5:7 and 5:8 resonant families, respectively.  In all 2D orbits $y_0=0$,
while the value of $p_{y0}$ is always found from the energy integral
(6).  The values of the initial conditions are given in the caption.
The integration time for all orbits, shown in Figures 6\,(a)--(h), is
100 time units.  It is remarkable that all the above resonances are
present in the simple 2D Hamiltonian (6).  The interaction of all these
resonances justifies the presence of the different sticky regions and
also the four different chaotic components.


\begin{figure}[!tH]
\begin{center}
\resizebox{0.80\hsize}{!}{\rotatebox{0}{\includegraphics*{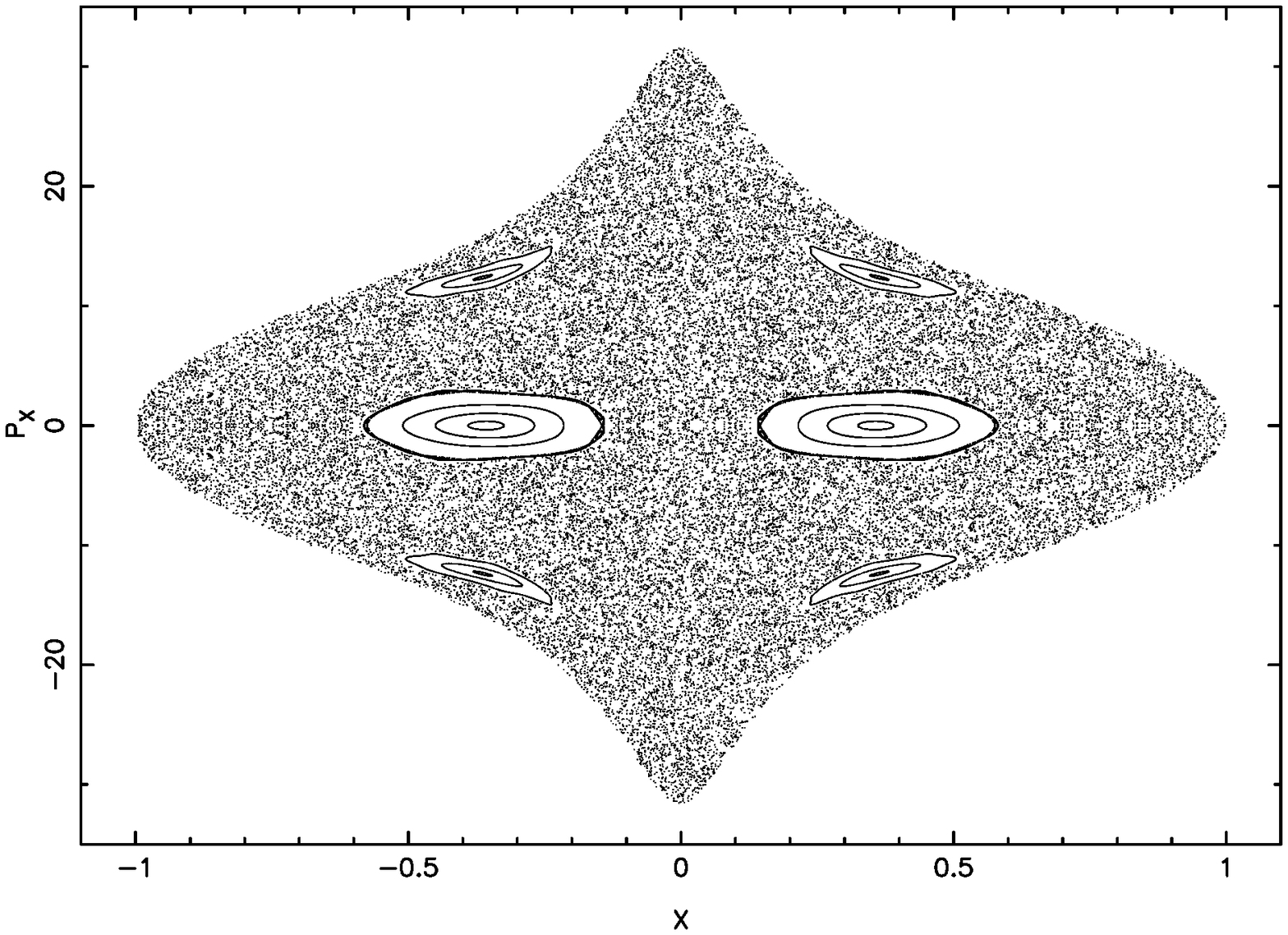}}}
\end{center}
\captionc{7}{Similar to Figure 1, but for $c_n=0.10$ and $h_2=0.25$.}
\end{figure}

Figure 7 is similar to Figure 1 but corresponds to $c_n=0.1$ and
$h_2=0.25$.  The only physical difference here is that we have a more
dense nucleus.  In this case, all the above chaotic components have
merged to produce one chaotic ocean.  There are only two sets of
islands.  The set of the invariant curves in the two large islands on
the $x$-axis is produced by quasi-periodic orbits typical for the 1:1
resonance, while the set of the four islands is produced by
quasi-periodic orbits typical for the 1:2 resonance.  Note that these
resonances are the basic ones of the system, since the 1:1 resonance
originates from the potential of the nucleus, and the 1:2 resonance
originates from the potential of the anisotropic harmonic oscillator.
There is only one sticky region left at the boundary between the set of
the large islands and the chaotic ocean.  Another interesting property
is the increase in the velocity near the central region of Figure 7.
Therefore, one can conclude that the nuclei with moderate densities
produce considerable sticky motions, different chaotic regions and a
large number of resonances.  On the other hand, the nuclei with high
densities produce large unified chaotic regions.  The degree of chaos in
this case is high, as the value of the LCE for a set of different
initial conditions on the chaotic ocean and a time period of $10^5$ time
units was found to tend to the value 4.25.  Furthermore, it is evident
that this chaotic motion happens at high velocities.

\sectionb{3}{THE STRUCTURE OF THE 3D DYNAMICAL SYSTEM}

In this Section, the results regarding the character of motion in the 3D
system will be presented.  For this purpose we use the 3D Hamiltonian
(2) and take $h=h_2$, i.e., the value of energy is equal to that in the
2D system.  The results obtained in the 2D system will be used to
investigate the motion in the 3D system.  We  consider orbits with a
starting point on the $x-p_x$ phase plane with an additional value of
$z=z_0$ and follow the evolution of 3D orbits.

As a first step in this investigation, we decided to compute the LCEs
for orbits with the initial conditions ($x_0$, $p_{x0}$, $z_0$), where
($x_0$, $p_{x0}$) is a point in each of the chaotic components of the 2D
system (see Figures 1 and 4).  The values of $y_0$ and $p_{z0}$ are
always taken equal to zero, while the value of $p_{y0}$ is always found
from the energy integral (2).  Figure 8 is similar to Figure 5 but
corresponds to the 3D system.  It is interesting to observe, that each
chaotic component has again a different value of LCE.  The LCE values
were found to be 0.4231, 0.0007 and 0.1923 in the chaotic components I,
II and III respectively.  In the chaotic sea the LCE has the value
0.4562.  Note that all the LCE values for the 3D system are smaller than
the corresponding values of the 2D system.  For the computation of the
LCE values of the chaotic components we used the value $z_0$ = 0.1.


\begin{figure}[!tH]
\centering
\resizebox{0.80\hsize}{!}{\rotatebox{0}{\includegraphics*{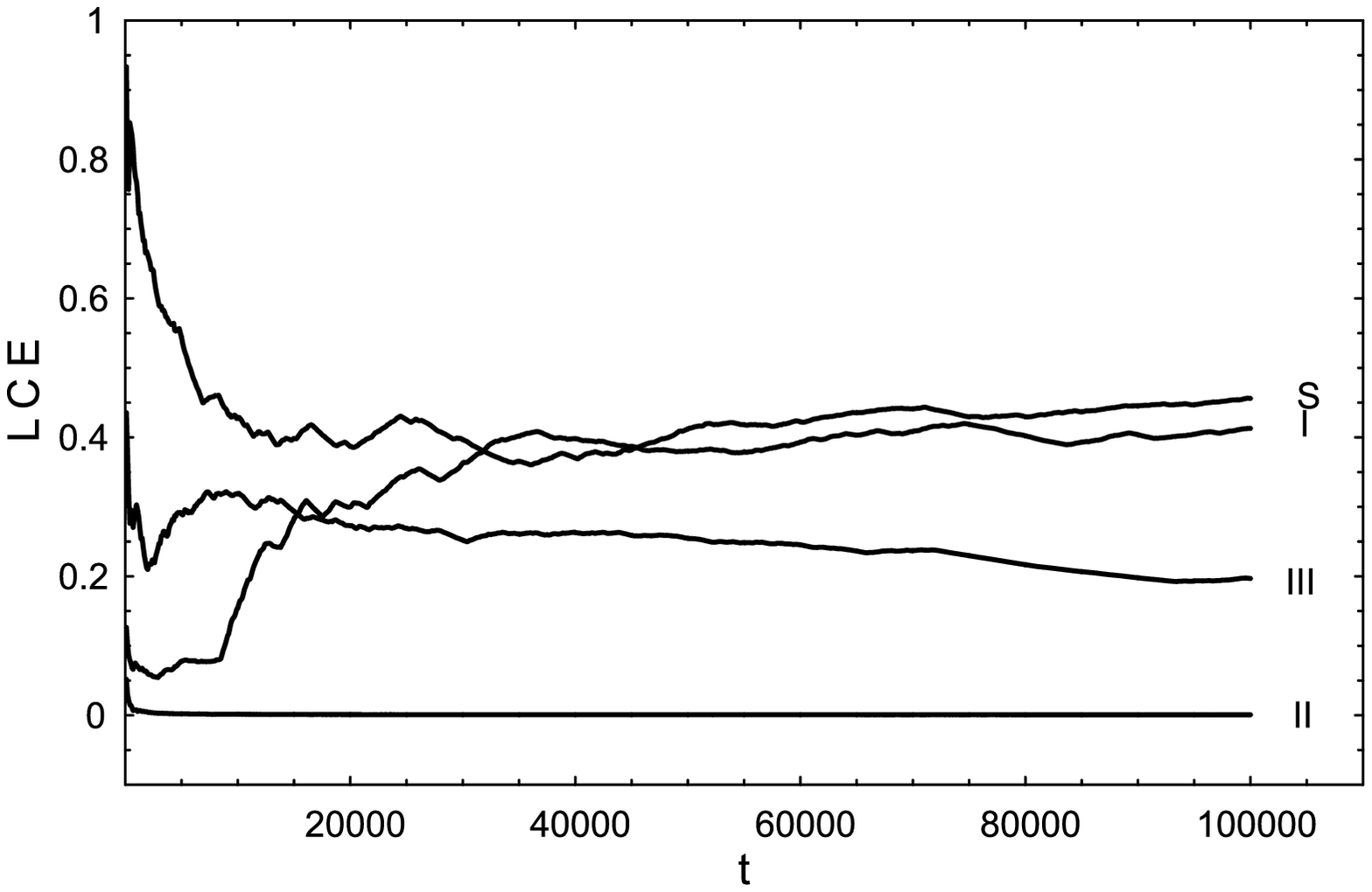}}}
\captionb{8}{Similar to Figure 5, but for the 3D model.}
\end{figure}

This strongly indicates, that the different chaotic components of the 3D
system do not merge to produce a single unified chaotic region.  This
result is compatible with the outcomes of Cincotta et al.  (2006), where
no evidence is found that in 3D systems with a divided phase space a
completely unified chaotic manifold actually exists.  On the other hand,
in the case of the dense nucleus, where we have only one chaotic
component, all tested 3D orbits with the initial conditions ($x_0$,
$p_{x0}$, $z_0$) (where ($x_0$, $p_{x0}$) is a point at the chaotic
ocean) lead to a common value of LCE close to 1.65.

Our next step is to see what happens to orbits with the initial
conditions
($x_0$, $p_{x0}$, $z_0$), where ($x_0$, $p_{x0}$) is a point in the
regular or chaotic regions of the phase plane of Figure 1. For this
purpose we introduce and use a new type of dynamical parameter -- the
$S(k)$ spectrum. The parameter $k_i$ is defined as
\begin{equation}
k_i=\frac{(x_i+z_i)-(p_{xi}+p_{zi})}{p_{yi}} \ ,
\end{equation}
where ($x_i$, $z_i$, $p_{xi}$, $p_{yi}$, $p_{zi}$) are the successive
values of ($x$, $z$, $p_x$, $p_y$, $p_z$) elements of the 3D orbit.  We
will call
as the dynamical spectrum of the parameter $k$ its distribution function
\begin{equation}
S(k)=\frac{\Delta N(k)}{N \Delta k} \ ,
\end{equation}
where $\Delta N(k)$ is the number of the parameters $k$ in the interval
$\left(k, k+\Delta k \right)$ after $N$ iterations.  By definition, the
$k$ parameter is based on a complicated combination of coordinates and
velocities
of the 3D orbit.  There are several reasons for introducing the $S(k)$
spectrum:  (1) it can identify 3D islandic motion, (2) it can help us to
understand the evolution of 3D sticky orbits, and (3) it is much faster
than other dynamical indicators, such as the LCE, SALI (Smaller
ALignment
Index, see Skokos 2001) or and GALI (Generalized ALignment Index, see
Skokos et al. 2007), which need time periods of the order of $10^5$ time
units to give reliable results.


\begin{figure*}[!tH]
\begin{center}
\resizebox{\hsize}{!}{\rotatebox{0}{\includegraphics*{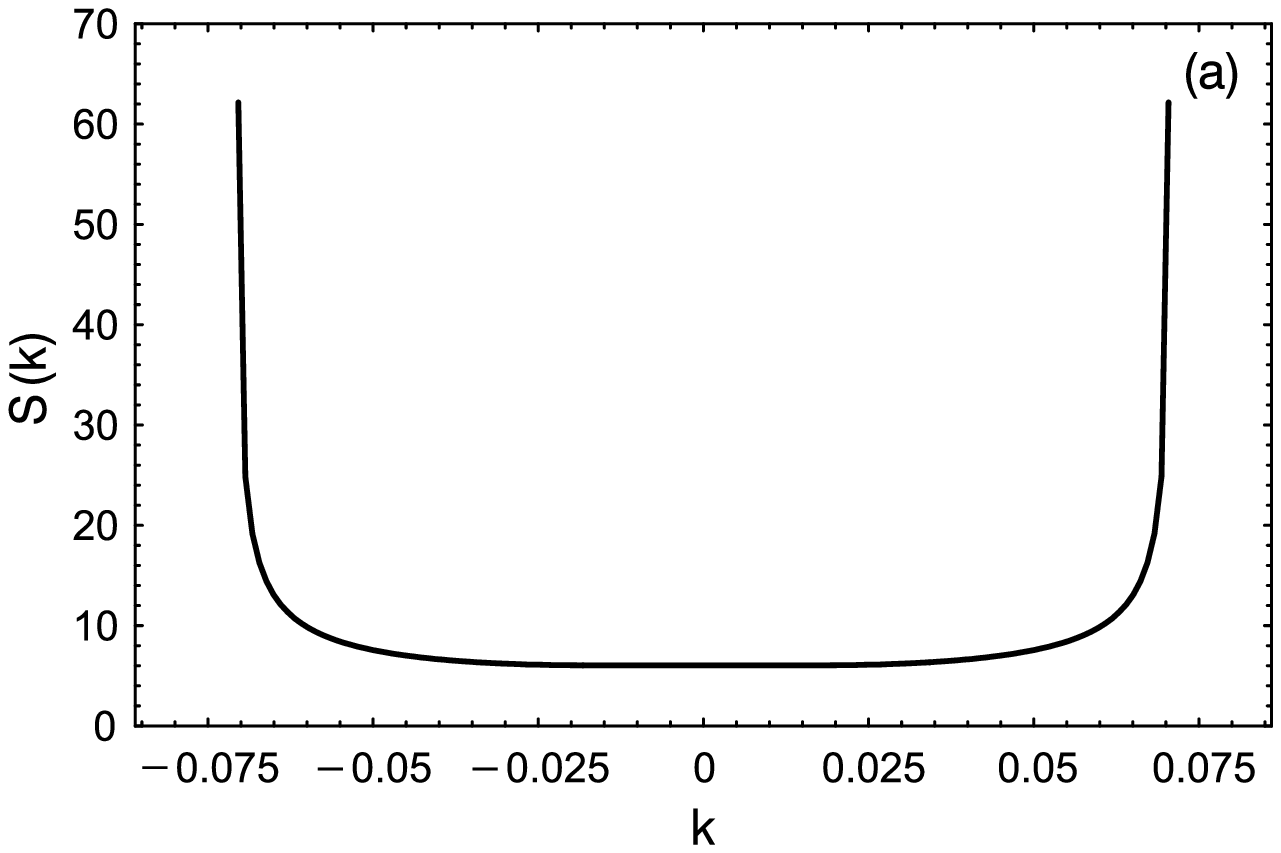}}\hspace{2cm}
                      \rotatebox{0}{\includegraphics*{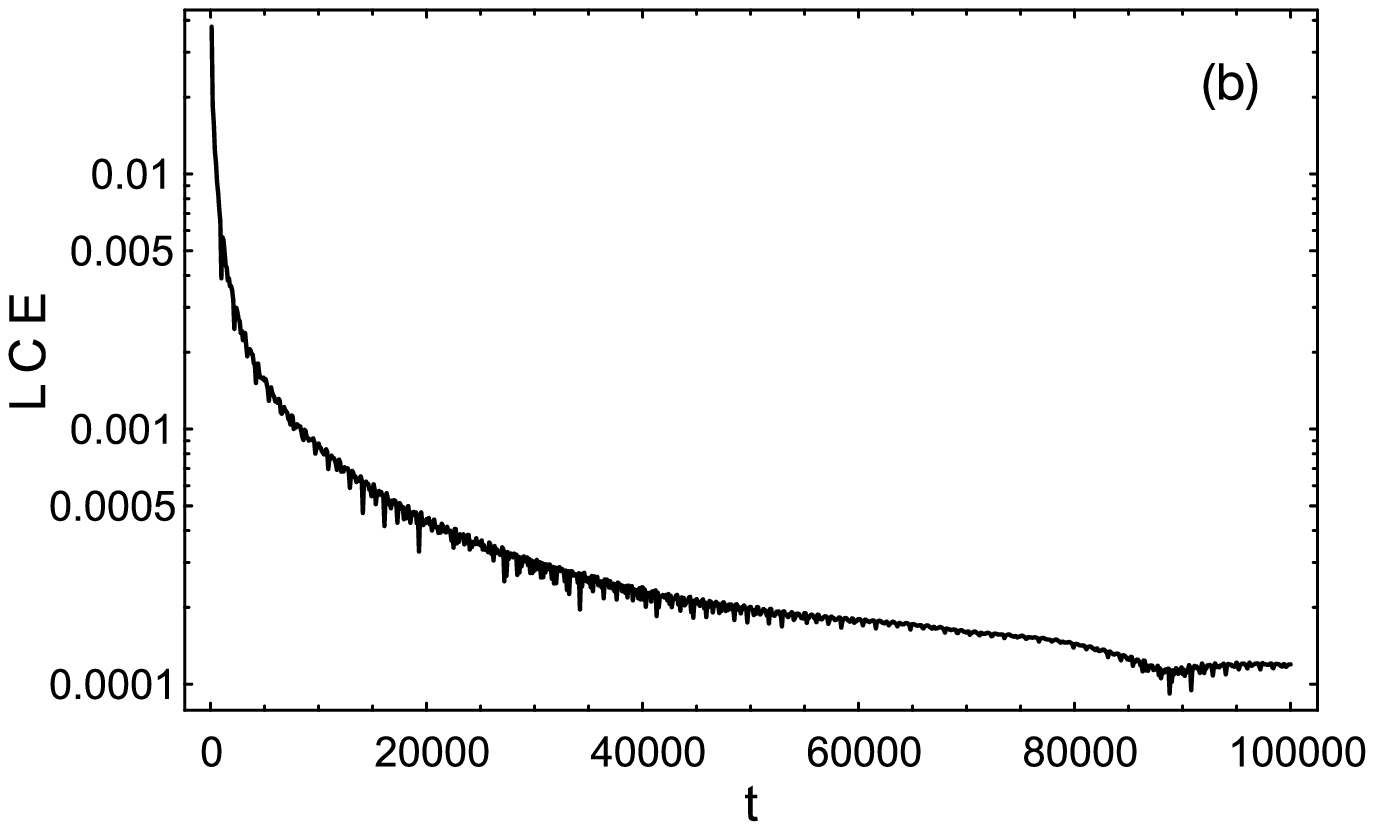}}}
\resizebox{\hsize}{!}{\rotatebox{0}{\includegraphics*{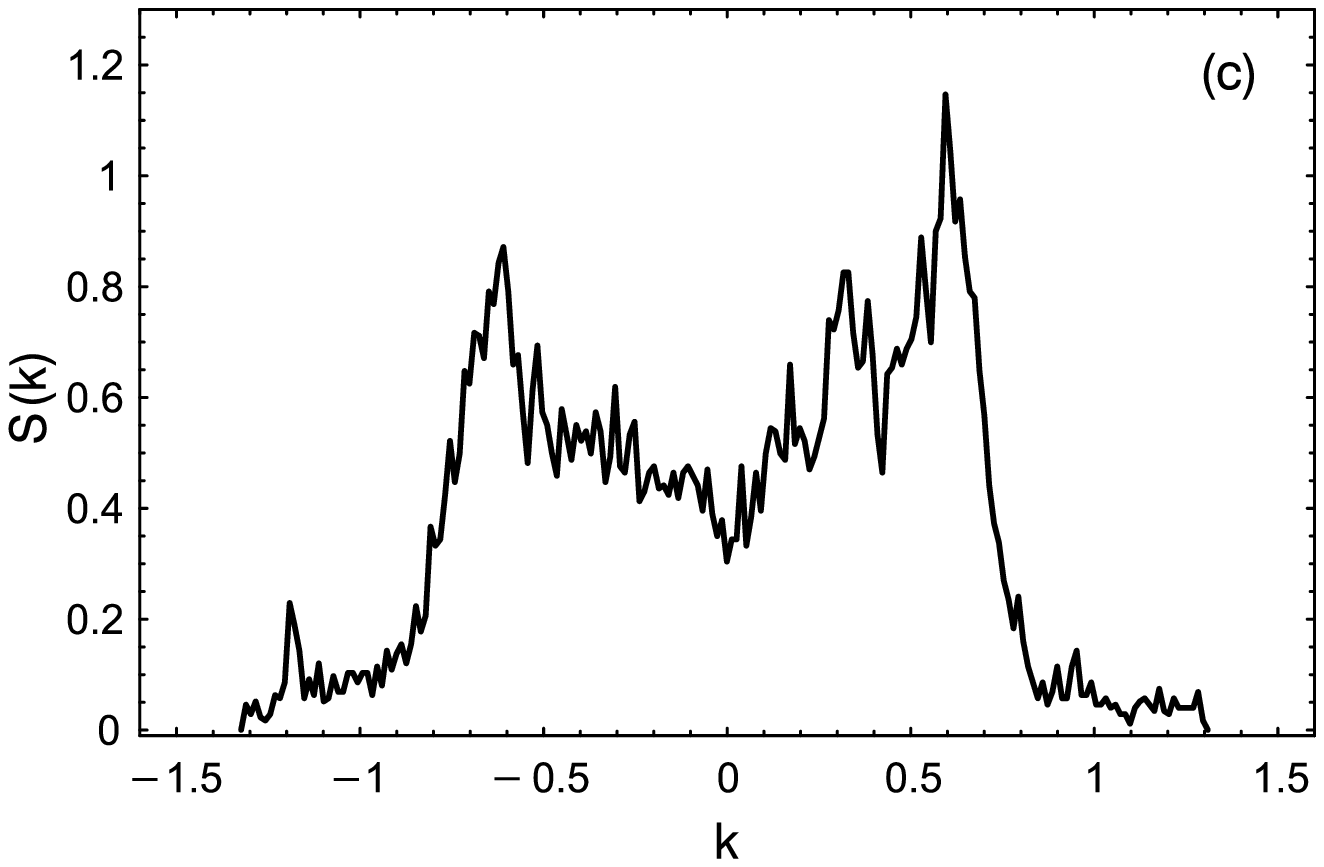}}\hspace{2cm}
                      \rotatebox{0}{\includegraphics*{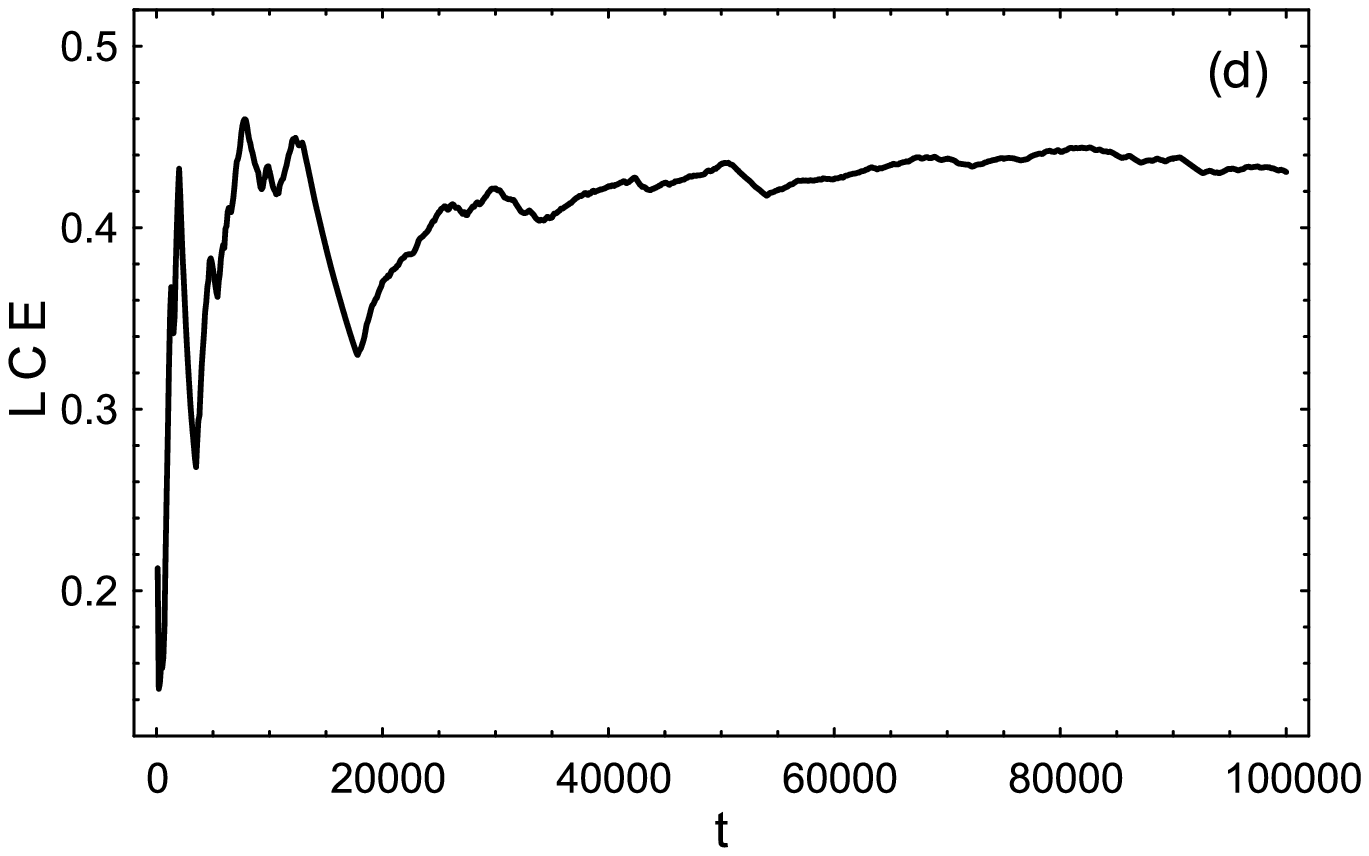}}}
\end{center}
\vskip 0.02cm
\captionb{9}{Panel (a): the $S(k)$ spectrum for a
3D regular orbit, panel (b): the corresponding LCE
of a 3D regular orbit, panel (c): the $S(k)$ spectrum for
a 3D chaotic orbit and panel (d): the corresponding
LCE of a 3D chaotic orbit.}
\end{figure*}

Figure 9a shows the $S(k)$ spectrum for a 3D regular orbit, with the
initial conditions:  $x_0=0.24$, $p_{x0}=p_{z0}=0$, $z_0=0.001$, while
the value of $p_{y0}$ is found from the energy integral (2).  As
expected, a well defined $U$-type spectrum is observed, indicating a
regular motion.  The corresponding LCE for a regular 3D orbit and a
time period of $10^5$ time units is given in Figure 9b.  Figure 9c
shows the $S(k)$ spectrum for a 3D chaotic orbit, with the initial
conditions:  $x_0=0.76$, $p_{x0}=1.4$, $p_{z0}=0$, $z_0=0.3$, and the
value of $p_{y0}$ is found from the energy integral (2).  Here, we can
see a complicated and asymmetric spectrum, with a lot of large and small
peaks.  This is a typical $S(k)$ spectrum of a chaotic orbit.  The
corresponding LCE for a 3D chaotic orbit and a time period of $10^5$
time units is given in Figure 9d.  The integration time for both
spectra shown in Figures 9a and 9c is 3000 time units.

Let us now start with the orbits with the initial conditions
($x_0$, $p_{x0}$) in the islands of Figure 2a.  Using the $S(k)$
spectrum it was found that for all ($x_0$, $p_{x0}$) in the regions of
Figure 2a, when $z_0 \leq 0.0065$, the motion is regular, while for all
larger permitted values of $z_0$ the motion becomes chaotic.


\begin{figure*}[!tH]
\begin{center}
\resizebox{\hsize}{!}{\rotatebox{0}{\includegraphics*{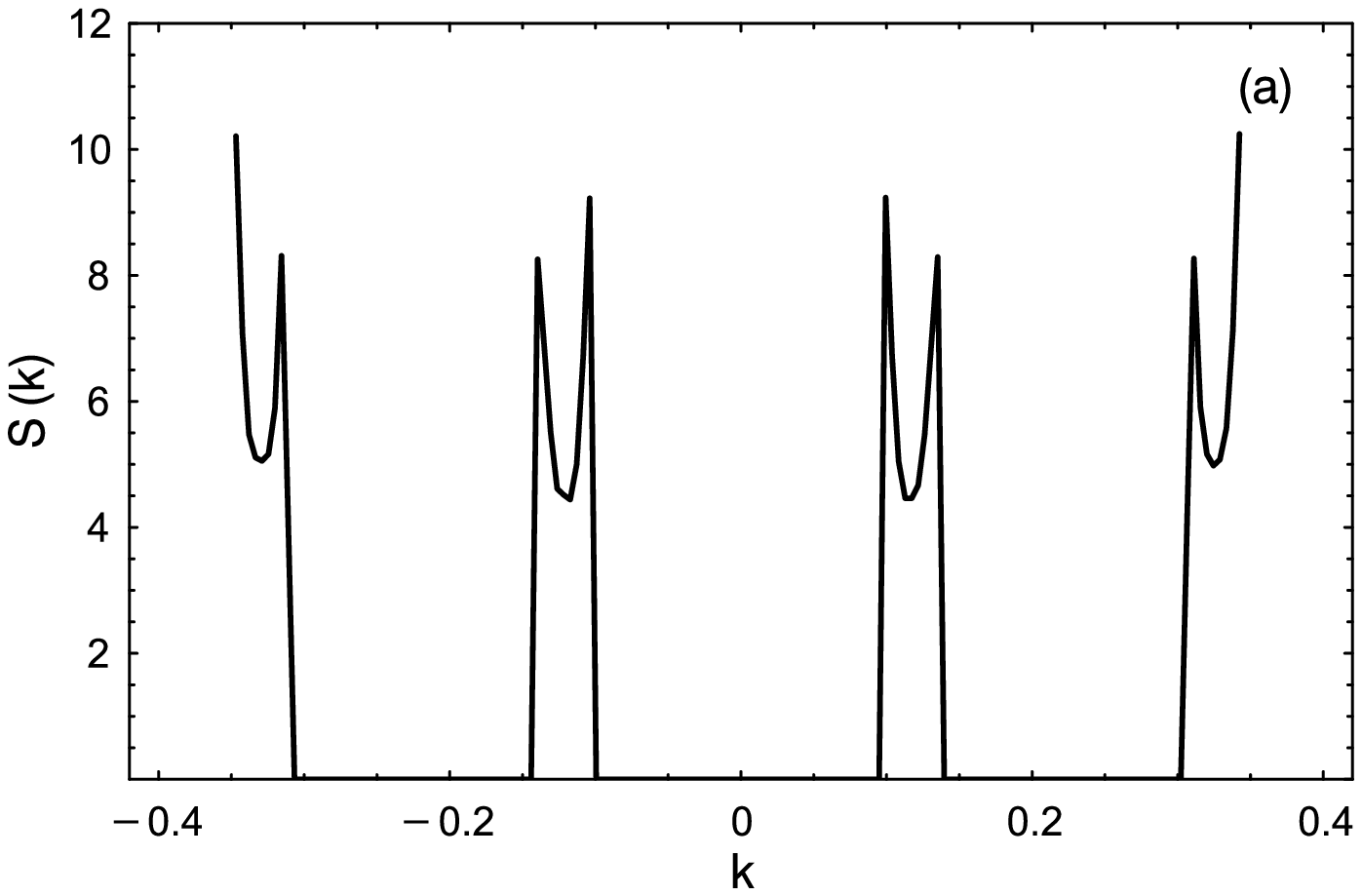}}\hspace{2cm}
                      \rotatebox{0}{\includegraphics*{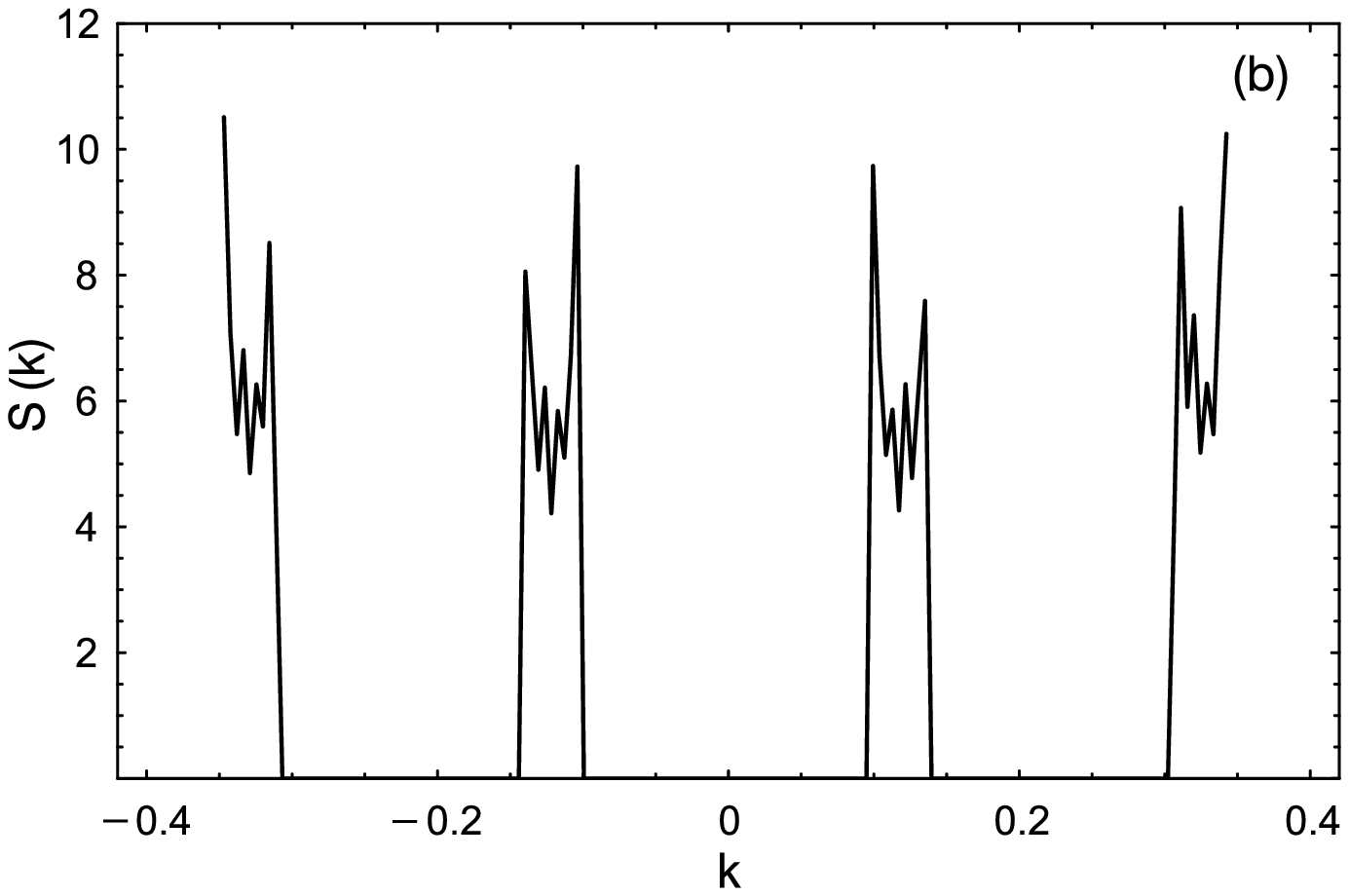}}}
\resizebox{\hsize}{!}{\rotatebox{0}{\includegraphics*{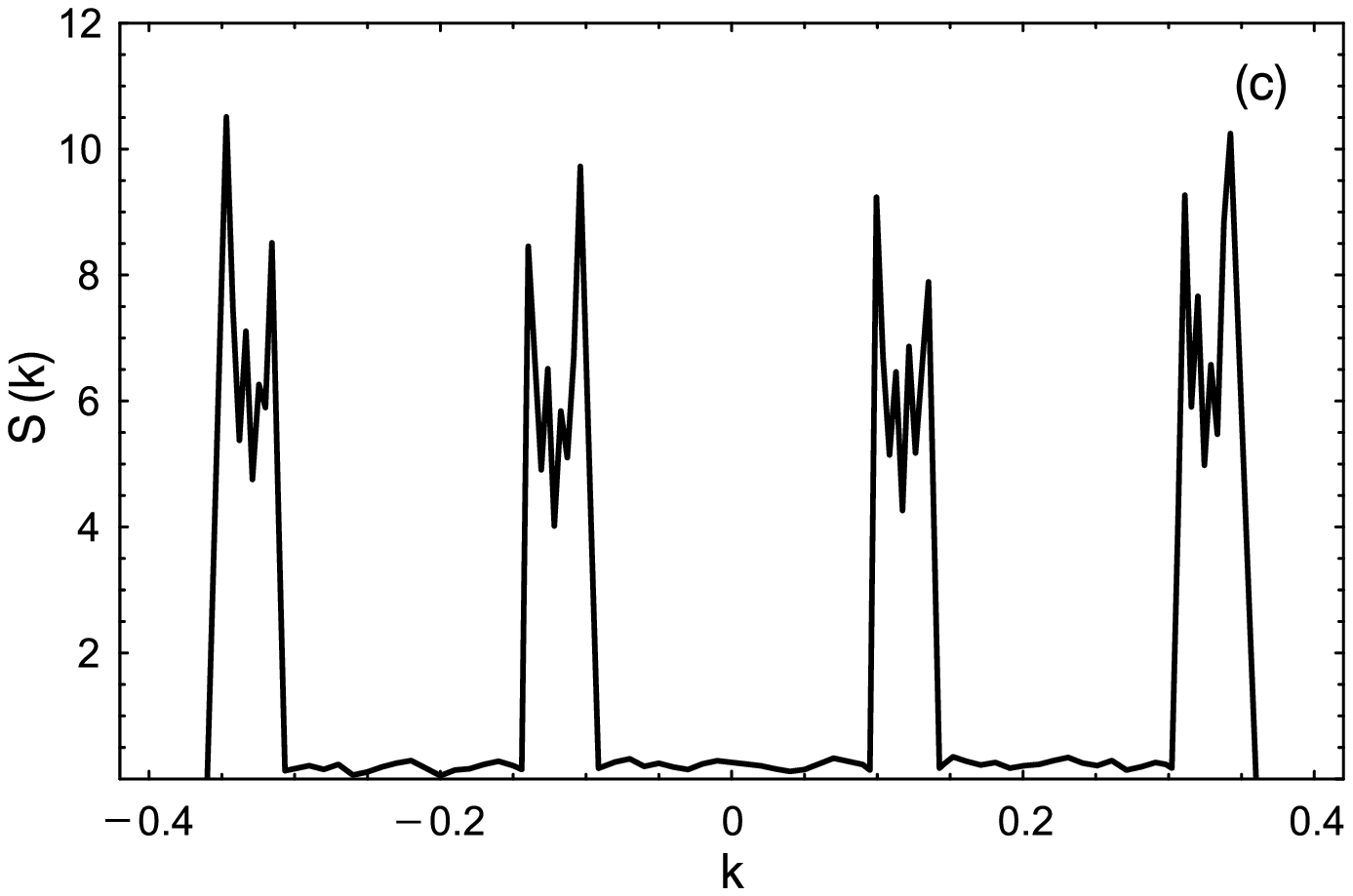}}\hspace{2cm}
                      \rotatebox{0}{\includegraphics*{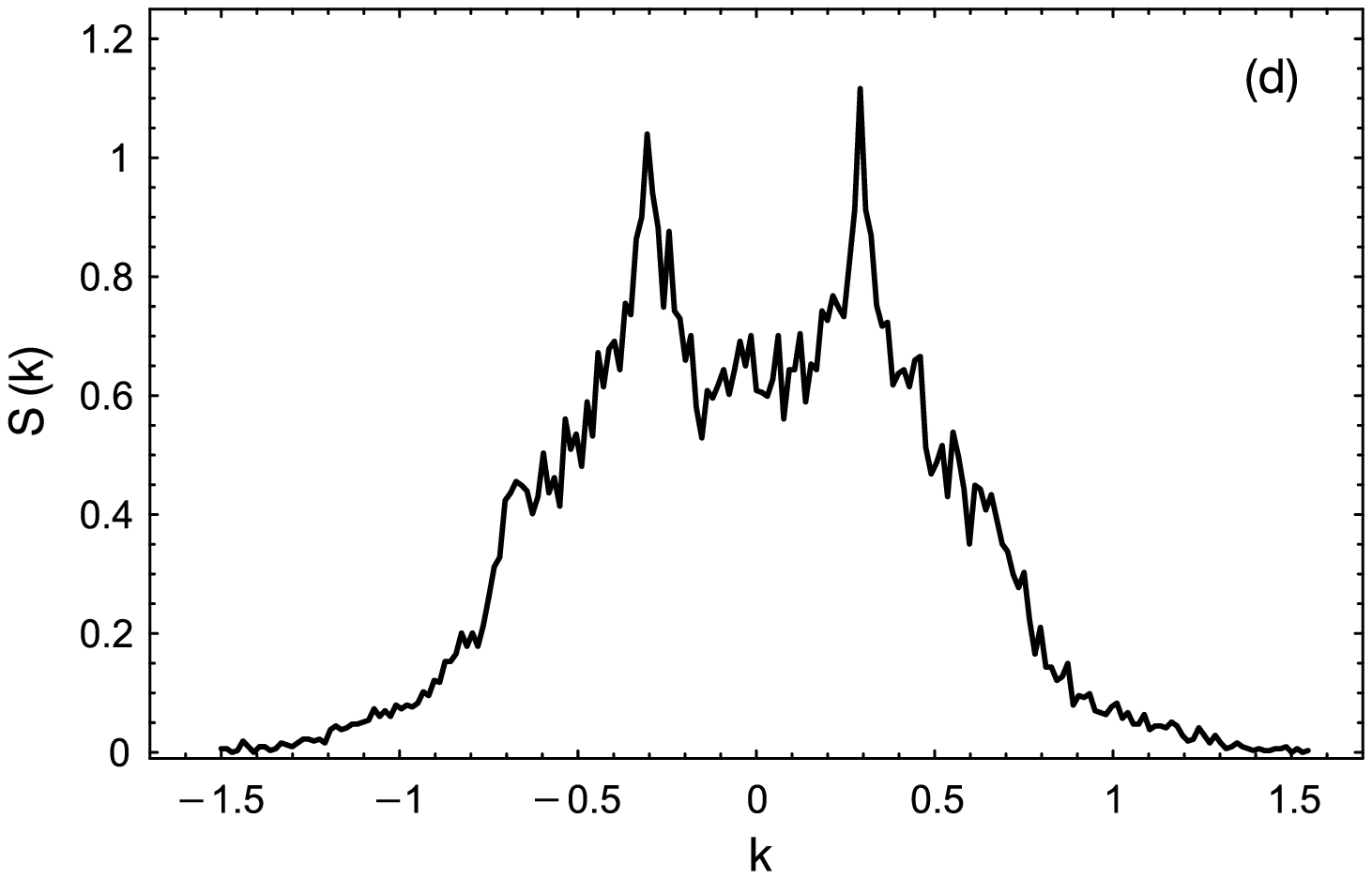}}}
\end{center}
\vskip 0.02cm
\captionb{10}{Panel (a): the $S(k)$ spectrum of a 3:4
resonant 3D orbit. Panels (b)--(d): the evolution of the $S(k)$
spectrum of a sticky 3D orbit. See the text for details.}
\vspace{5mm}
\end{figure*}


\begin{figure*}[!tH]
\begin{center}
\resizebox{\hsize}{!}{\rotatebox{0}{\includegraphics*{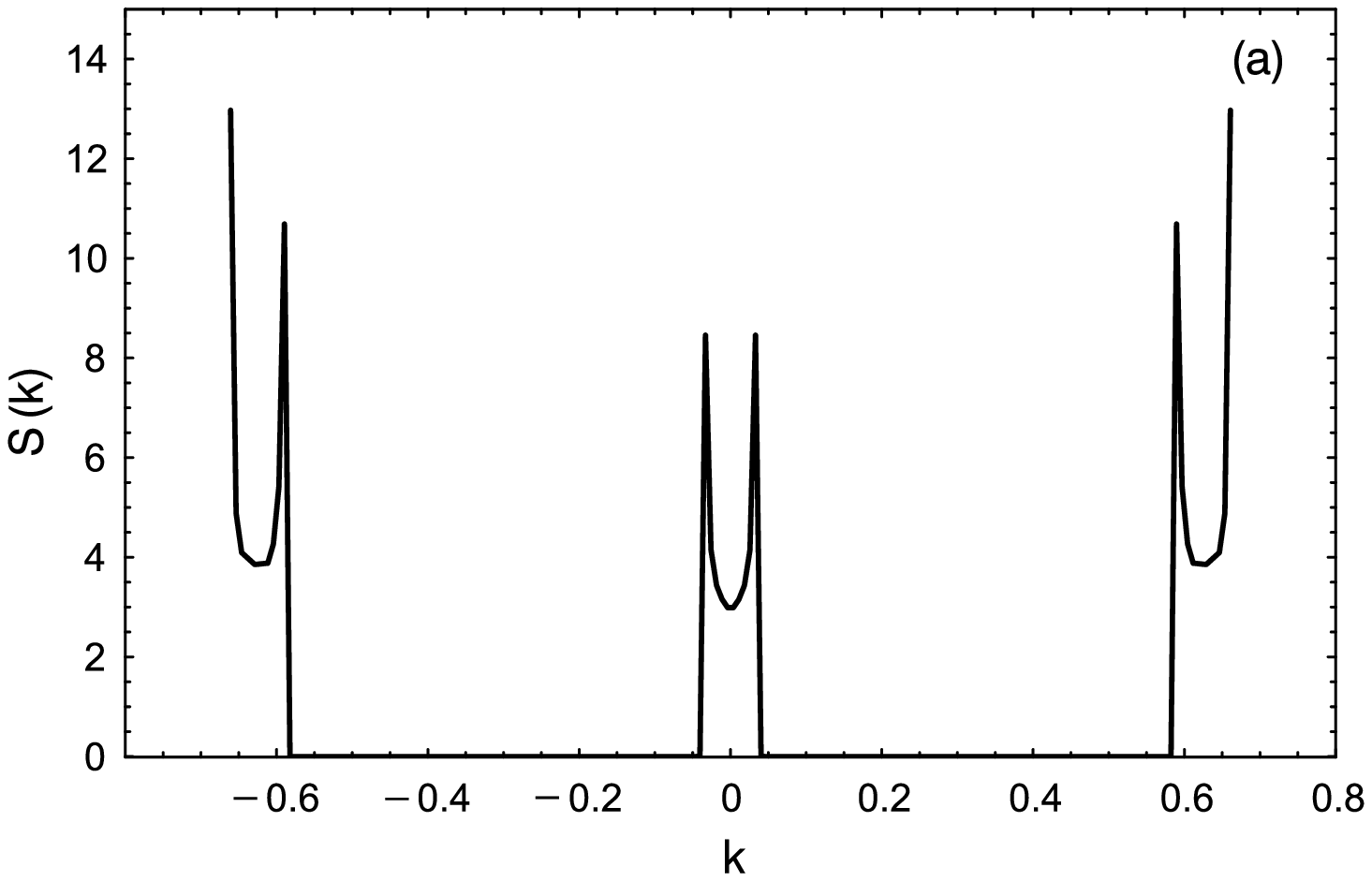}}\hspace{2cm}
                      \rotatebox{0}{\includegraphics*{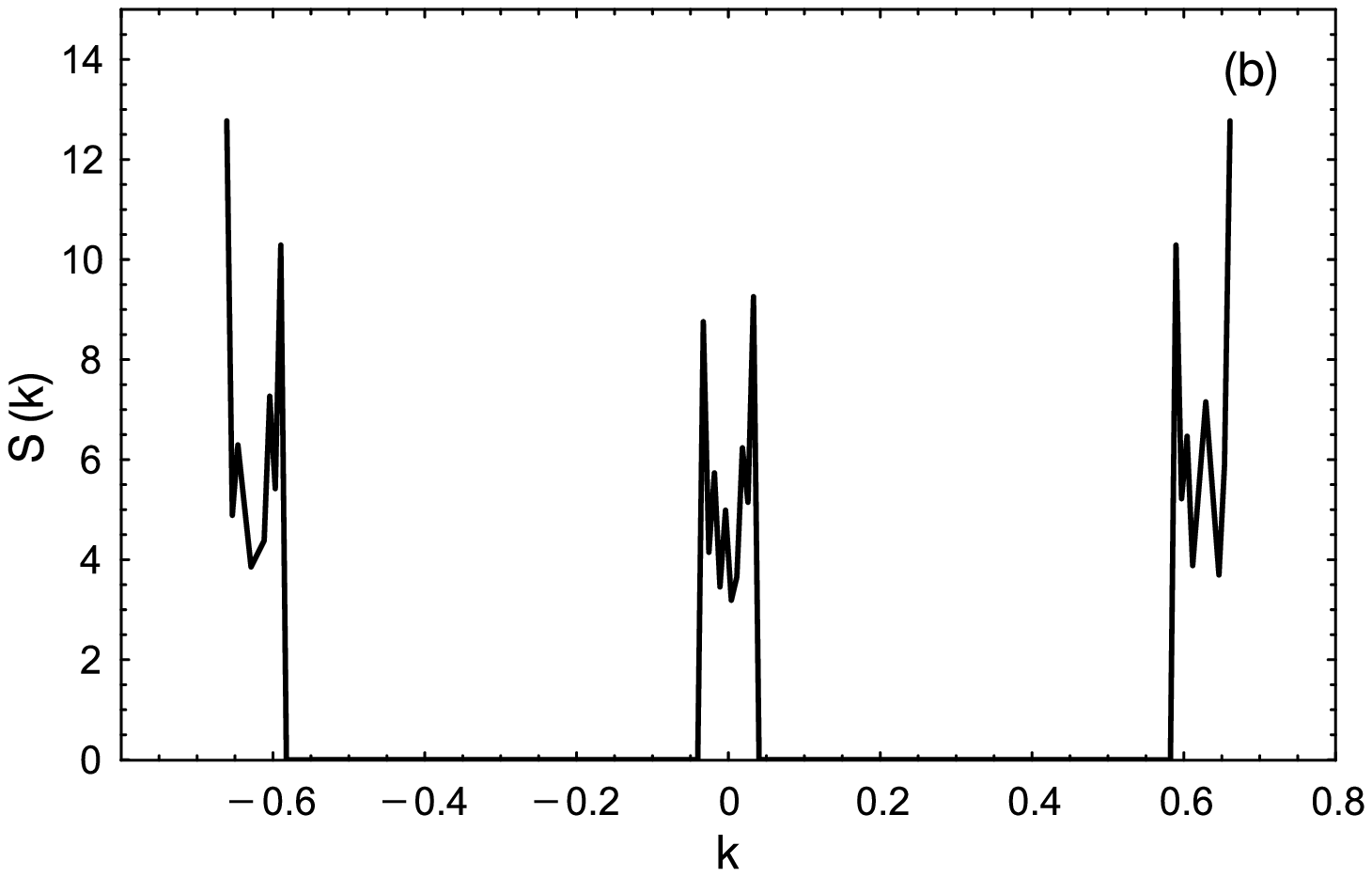}}}
\resizebox{\hsize}{!}{\rotatebox{0}{\includegraphics*{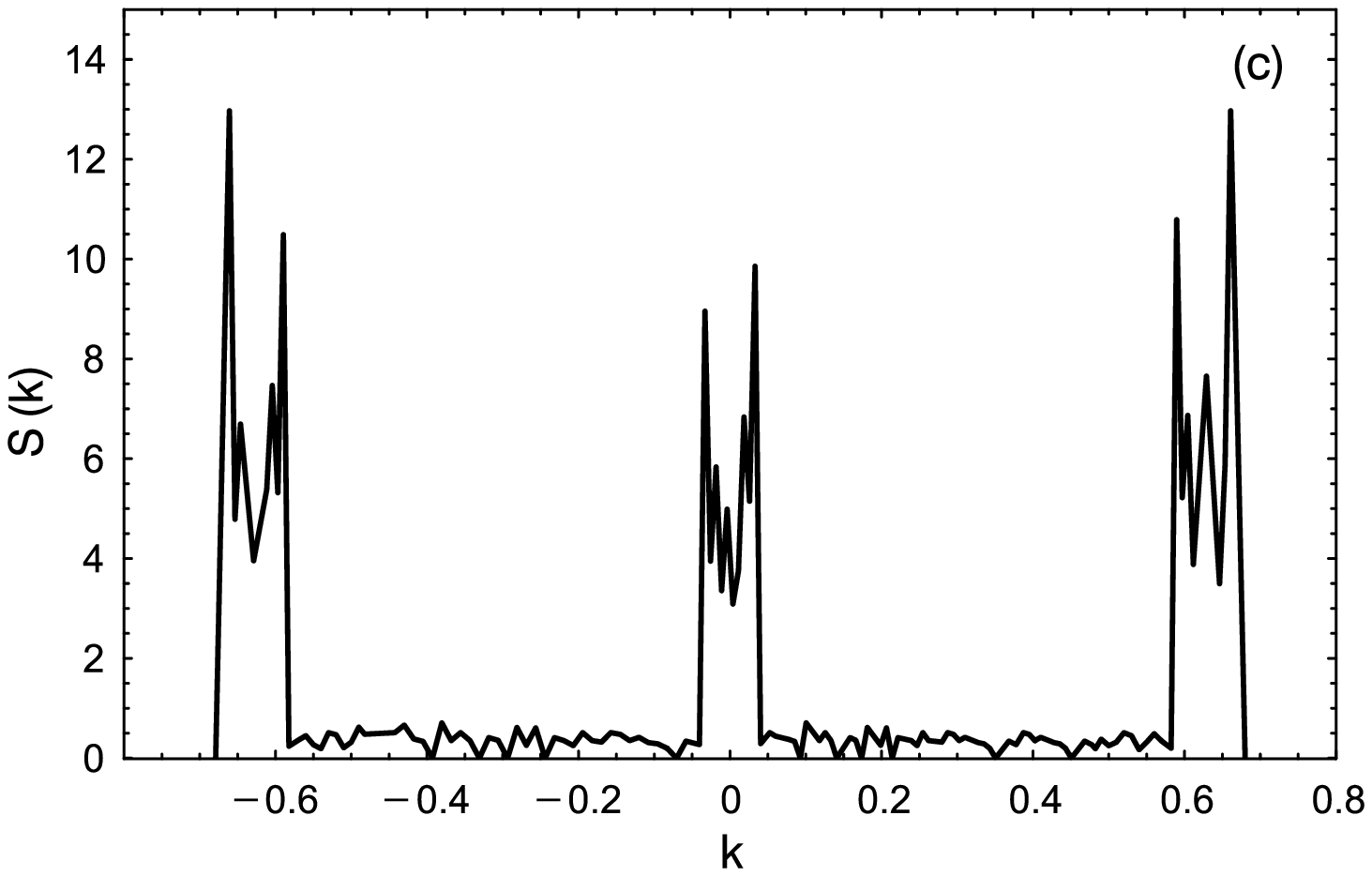}}\hspace{2cm}
                      \rotatebox{0}{\includegraphics*{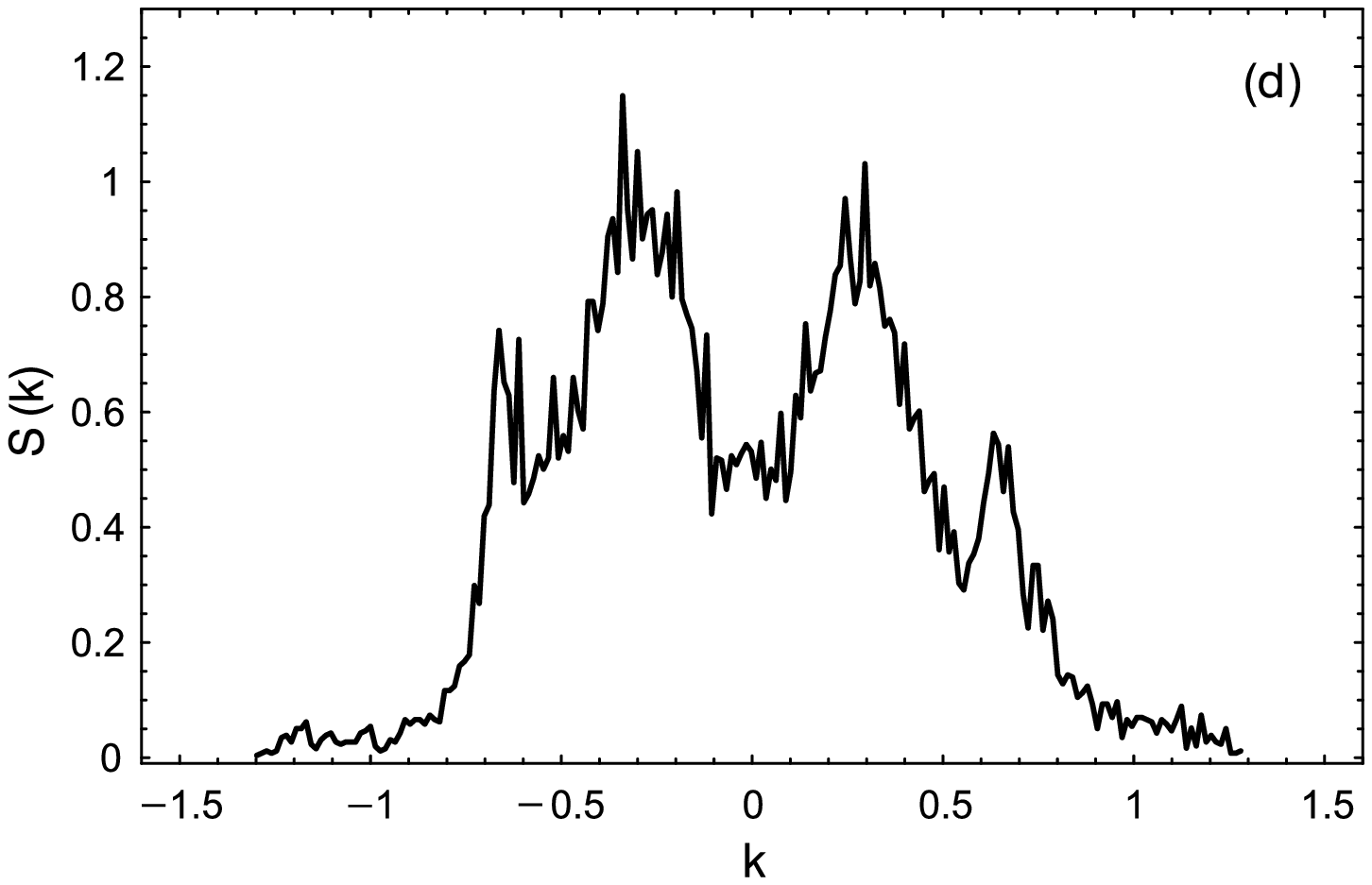}}}
\end{center}
\vskip 0.02cm
\captionb{11}{Panel (a): the $S(k)$ spectrum of a 2:3
resonant 3D orbit. Panels (b)--(d): the evolution of the $S(k)$
spectrum of a sticky 3D orbit. See the text for details.}
\end{figure*}

Interesting results are obtained for the initial conditions
($x_0$, $p_{x0}$, $z_0$), where ($x_0$, $p_{x0}$) is a point in the
regular region inside one of the four islands in the phase plane of
Figure 2b.  Remember that in Figure 2b we see eight islands produced by
the two identical orbits traversed in opposite directions.  Each of
these orbits produces four islands.  Figure 10\,(a)--(d) shows the
$S(k)$ spectrum of two such orbits.  Figure 10a shows the $S(k)$
spectrum for an orbit producing the set of four islands shown in Figure
2b.  The initial conditions are:  $x_0=0.09$, $p_{x0}=7.5$, $z_0=0.01$,
$y_0=p_{z0}=0$.  As expected, we observe four well defined $U$-type
spectra.  The motion is regular.  Figure 10b shows the $S(k)$ spectrum
for the orbit starting near the above described regular orbit.  The
initial conditions are:  $x_0=0.09$, $p_{x0}=7.5$, $z_0=0.05$,
$y_0=p_{z0}=0$.  Here we see again the four spectra, each one
corresponding to an island.  Figure 10b exhibits some additional
small peaks which indicate a 3D sticky motion.

Let us now look at the evolution of a 3D sticky orbit using the $S(k)$
spectrum.  The sticky period is about 1200 time units.  Figure 10c shows
the $S(k)$ spectrum of a sticky orbit, about 100 time units after the
test particle (star) has left the sticky region.  Here, the four spectra
have been joined together producing a single spectrum.  This spectrum
has the characteristics of a chaotic spectrum and this strongly suggests
that the test particle (star) has left the sticky region and entered to
the chaotic component.  The shape of the $S(k)$ spectrum, after the test
particle (star) has traveled 5000 time units in the chaotic component,
is shown in Figure 10d.  Our numerical experiments give similar results
for other initial conditions ($x_0$, $p_{x0}$, $z_0$), where ($x_0$,
$p_{x0}$) is a point inside one of the four islands of invariant curves
shown in Figure 2b.  Therefore, we come to the conclusion that the
orbits which start inside the set of the four islands, produced by the
3:4 resonance and having $z_0 \leq 0.038$, are regular while for larger
values of $z_0$ the orbits become sticky, with a sticky period of about
1200 time units.


\begin{figure*}[!tH]
\begin{center}
\resizebox{0.7\hsize}{!}{\rotatebox{0}{\includegraphics*{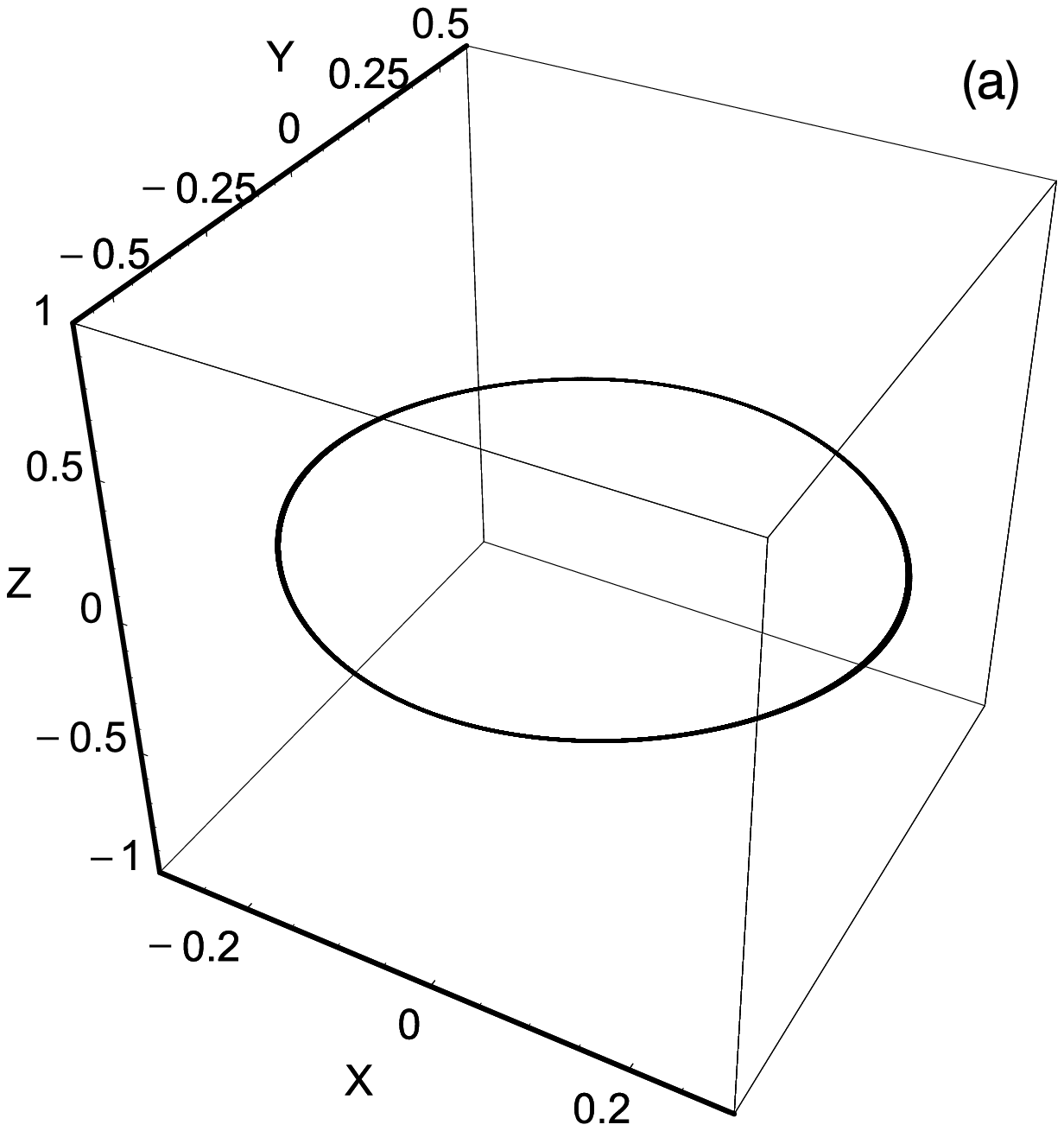}}\hspace{6cm}
                         \rotatebox{0}{\includegraphics*{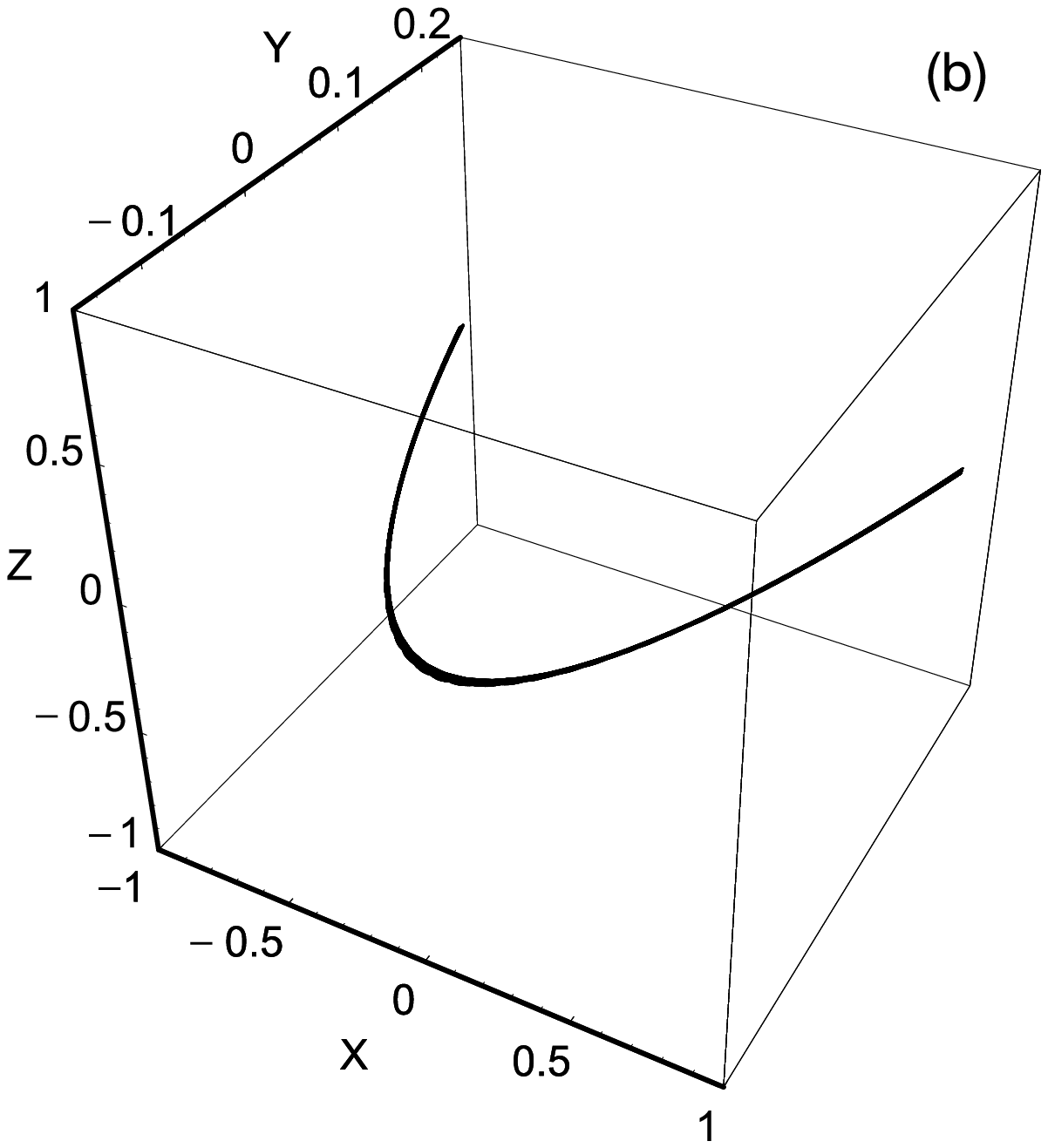}}}
\resizebox{0.7\hsize}{!}{\rotatebox{0}{\includegraphics*{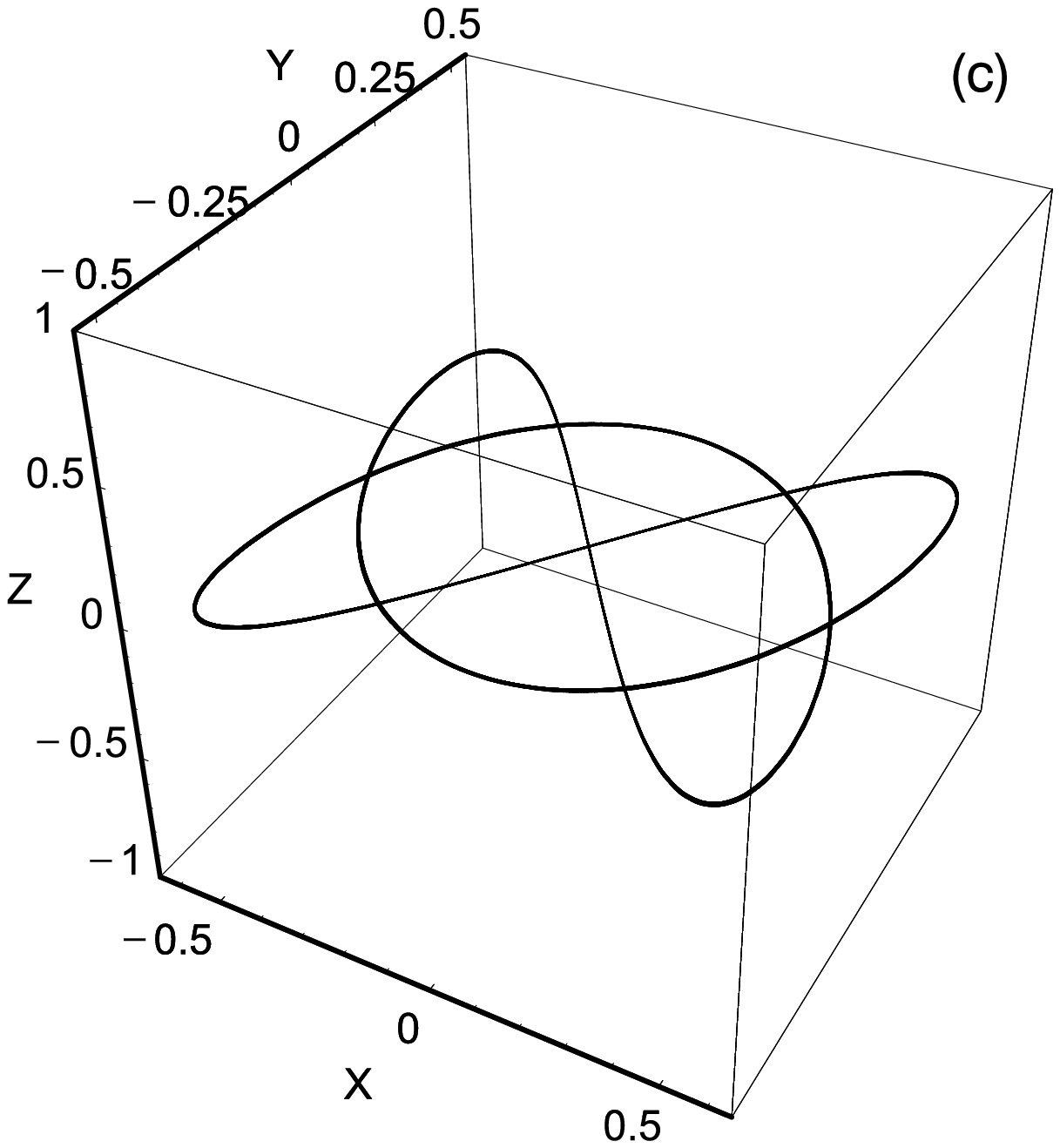}}\hspace{6cm}
                         \rotatebox{0}{\includegraphics*{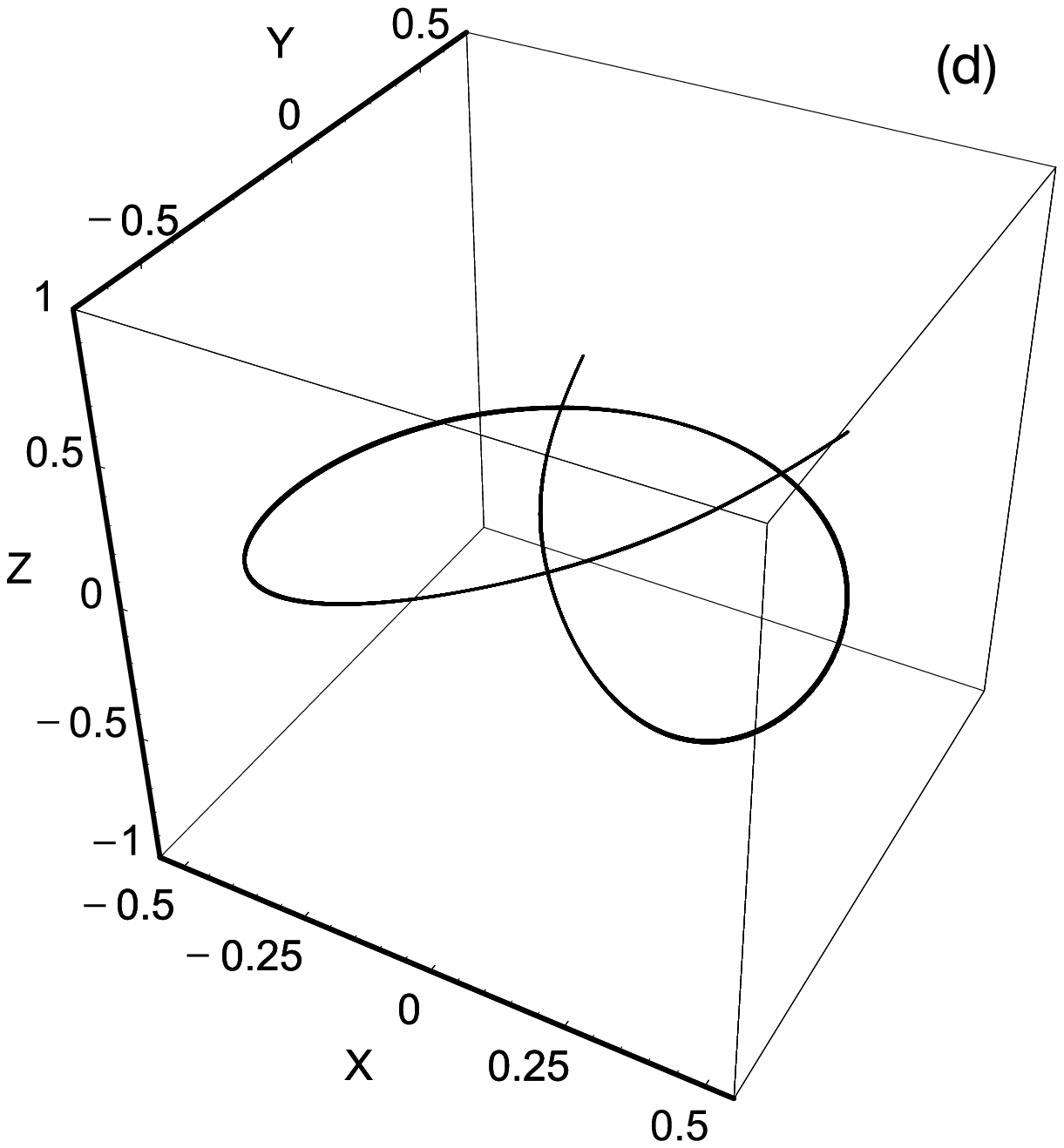}}}
\resizebox{0.7\hsize}{!}{\rotatebox{0}{\includegraphics*{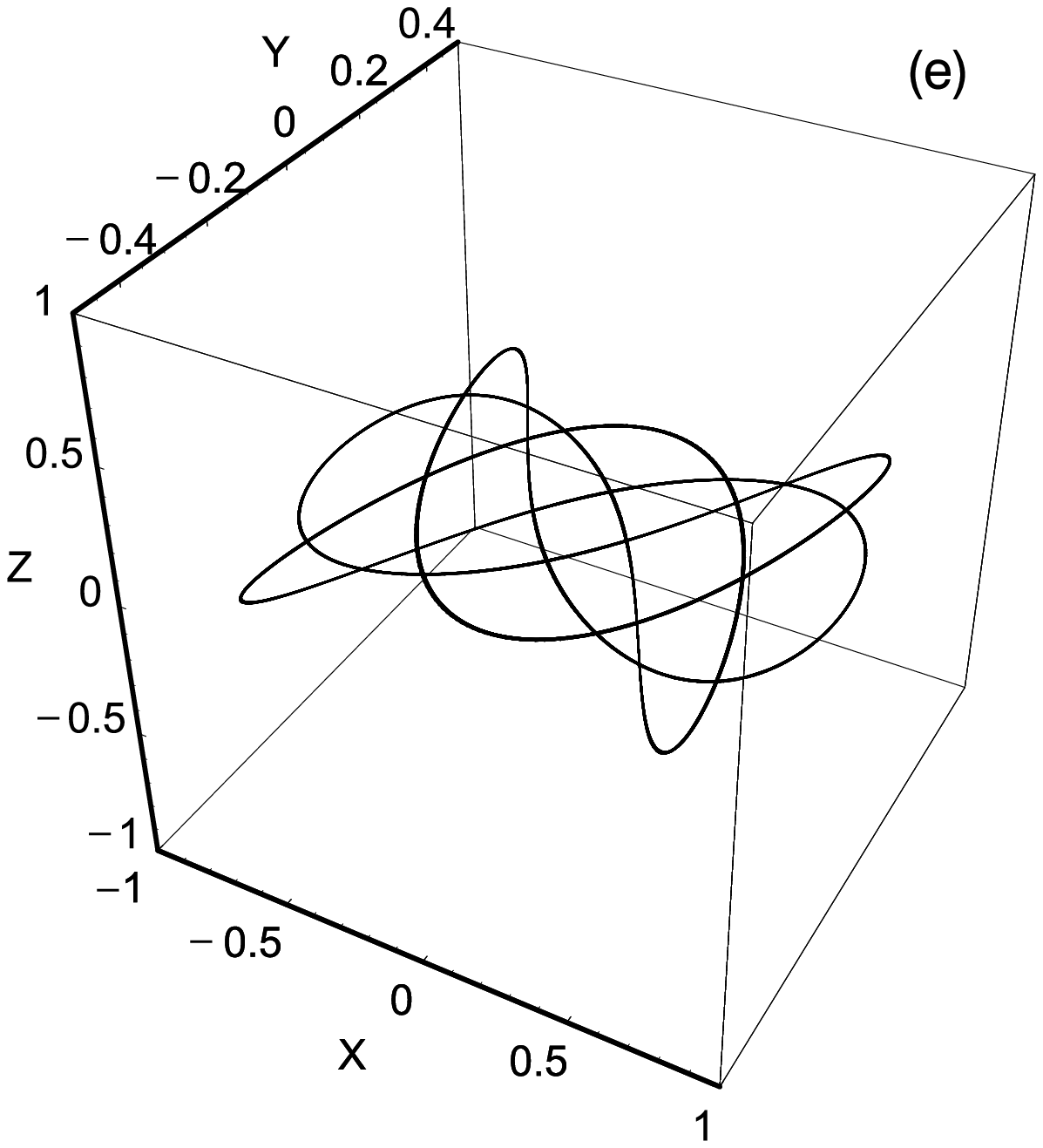}}\hspace{6cm}
                         \rotatebox{0}{\includegraphics*{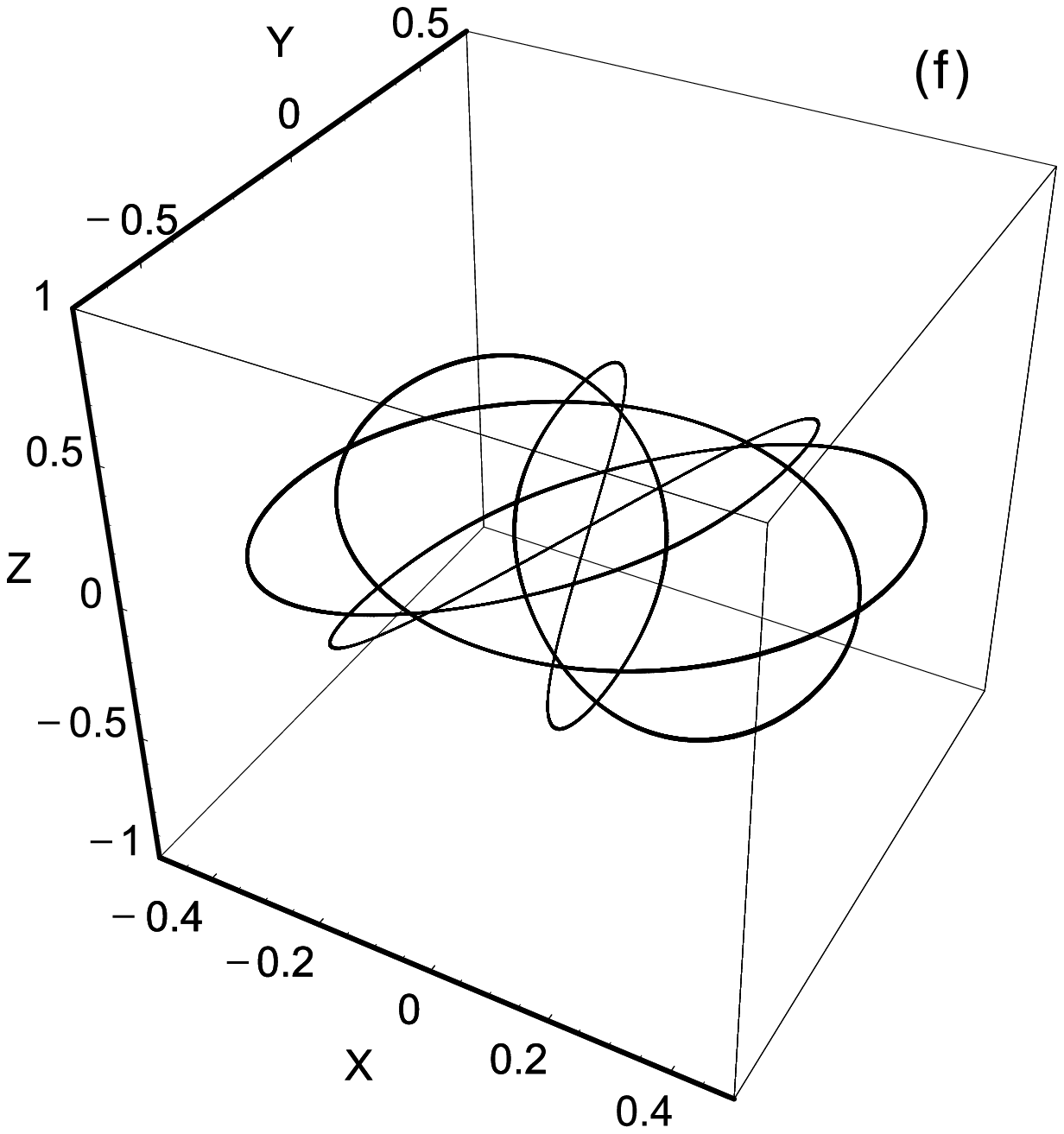}}}
\resizebox{0.7\hsize}{!}{\rotatebox{0}{\includegraphics*{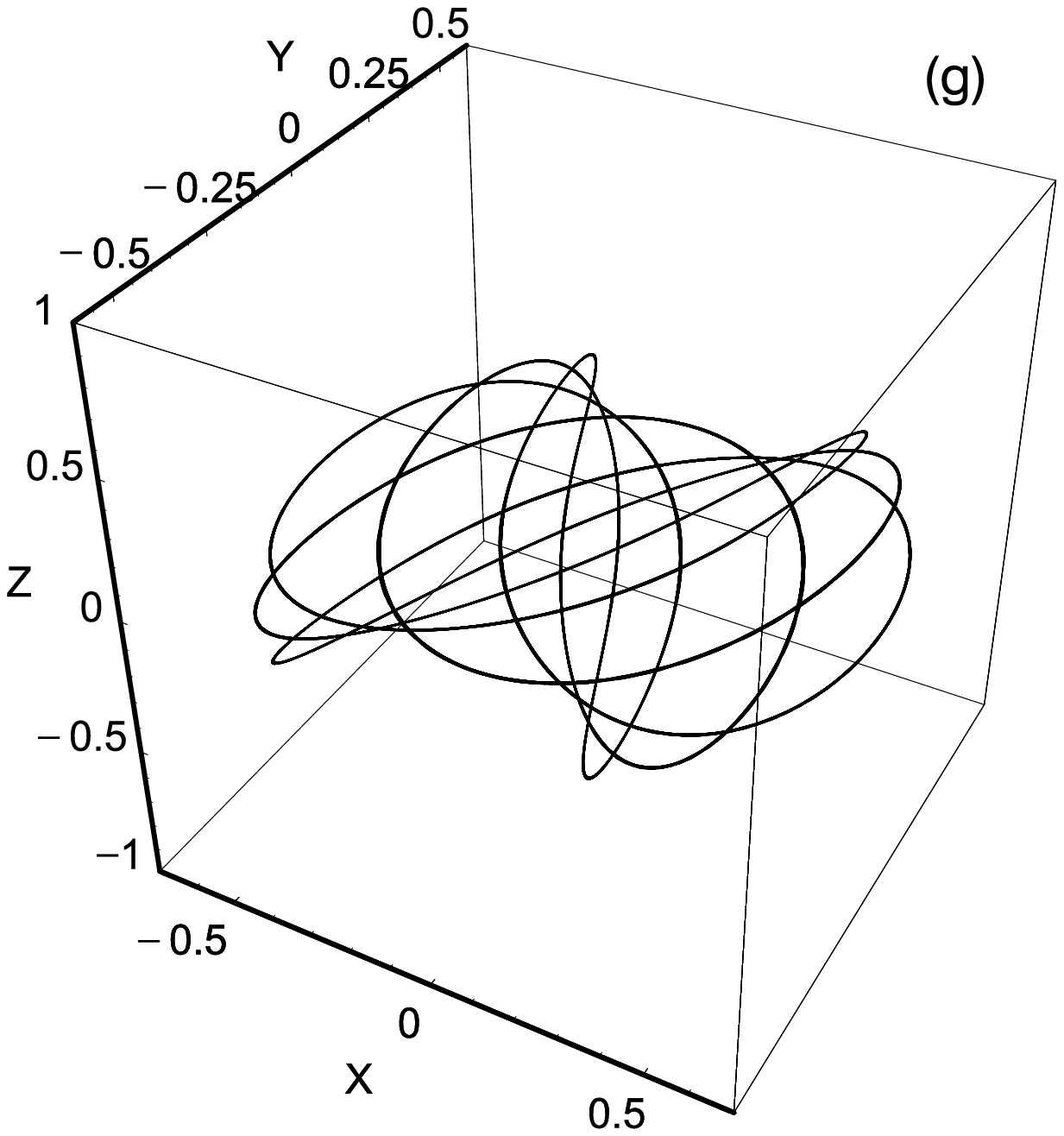}}\hspace{6cm}
                         \rotatebox{0}{\includegraphics*{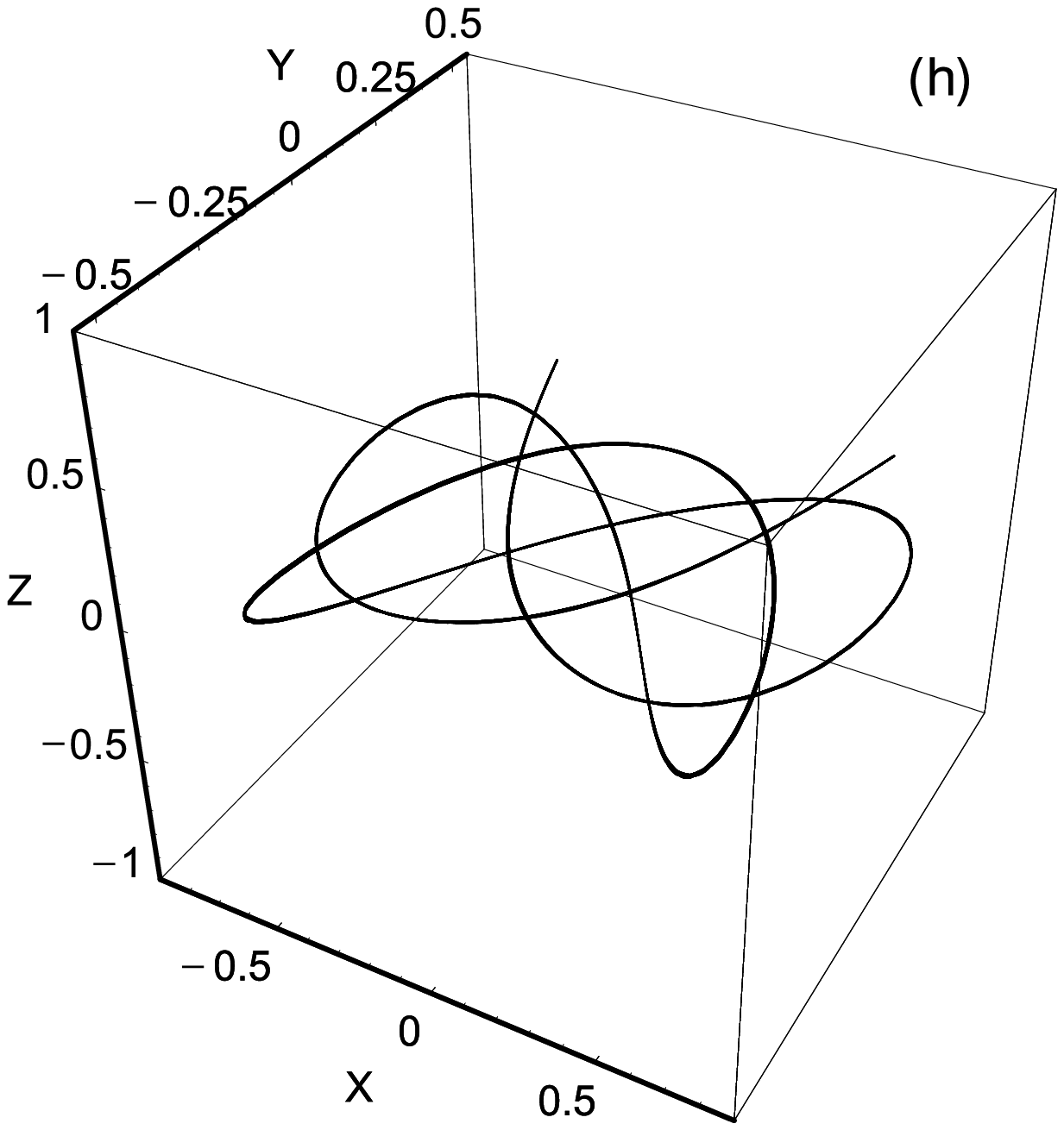}}}
\end{center}
\vskip 0.01cm
\captionb{12}{Panels (a)--(h): representative periodic orbits in the 3D
dynamical system. See the text for details.}
\end{figure*}

Things are similar for the orbits with the initial conditions ($x_0$,
$p_{x0}$, $z_0$), where ($x_0$, $p_{x0}$) is a point in the regular
region inside one of the three islands in the phase plane of Figure 2c.
Remember again that in Figure 2c we observe six islands produced by two
identical orbits traversed in opposite directions.  Each of these orbits
produces three islands.  Figure 11a shows the $S(k)$ spectrum of an
orbit with the initial conditions:  $x_0=0$, $p_{x0}=12.8$, $z_0=0.01$,
$y_0=p_{z0}=0$, and the value of $p_{y0}$  found from the energy
integral (2).  The values of all other parameters are as in Figure 1.
Here we observe three well defined $U$-type spectra indicating a regular
motion.  Note that the number of spectra is equal to the number of
islands.  Figure 11b shows the $S(k)$ spectrum of an orbit starting
near the above regular orbit.  The initial conditions are:  $x_0=0$,
$p_{x0}=12.8$, $z_0=0.05$, $y_0=p_{z0}=0$.  Here we observe three
different spectra with a number of large and small peaks.  This
indicates that we have a 3D sticky orbit, and the sticky region is
composed of three sticky islands.  The sticky period is about $T=1700$
time units.  Figure 11c shows the spectrum of the same orbit when
$T=1800$.  Here the three spectra have merged to produce a unified
spectrum.  This indicates that after the sticky period, the test
particle (star)
has gone to the 3D chaotic component.  In Figure 11d we see the spectrum
of the orbit when $T=7000$.  Here the spectrum of a chaotic orbit shows
a large number of small and large asymmetric peaks.  Our numerical
experiments give similar results for different initial conditions,
($x_0$, $p_{x0}$, $z_0$), where ($x_0$, $p_{x0}$) is a point inside one
of the three islands of invariant curves shown in Figure 2c.  Therefore,
once more we conclude that the orbits which start inside the set of the
three islands, produced by the 2:3 resonance and with $z_0 \leq 0.042$,
are regular, while for larger values of $z_0$ the orbits become sticky,
with a sticky period of about 1700 time units.  For each case of the
resonant orbits we have computed about 100 orbits with different initial
conditions in order to estimate the mean
value for each sticky period.

Figures 12\,(a)--(h) show eight typical periodic orbits in the 3D
dynamical system.  In all orbits the values of the initial conditions
($x_0$, $p_{x0}$) are the same as the 2D orbits, shown in Figures
6\,(a)--(h).  For all 3D orbits $y_0=0$, the value of $p_{y0}$ is found
from the energy integral (2) and the value of $z_0$ is 0.01.  Note that
all
3D periodic orbits shown in Figure 12\,(a)--(h) stay very close to the
galactic plane.  The integration time for all 3D periodic orbits, shown
in Figures 12\,(a)--(h), is 100 time units.

\sectionb{4}{DISCUSSION AND CONCLUSIONS}

In the present article we studied the orbital behavior in the central
parts of a triaxial elliptical galaxy hosting a dense and massive
nucleus. The main conclusions of our investigation are the following.

1. The orbital analysis of the Hamiltonian system of the two degrees of
freedom (2D) revealed a large variety of resonant orbits, different
chaotic components and several sticky regions. The different chaotic
components of the 2D system do not merge to produce a unified chaotic
region, even for time intervals much larger than the age of the
galaxy.

2. The results of the Hamiltonian system of the three degrees of freedom
(3D) are also of particular interest. Once more, the four different
chaotic components of the 3D dynamical system continue to exist
separately for vast time intervals, leading to the conclusion that the
chaotic phase space of the 3D system remains divided.

The general conclusion -- our model indicates that in the central region
of triaxial elliptical galaxies the 3D motion is complicated, displaying
several families of resonant orbits, different chaotic components and
remarkable sticky regions, while only a small fraction of orbits is
regular.  This interesting galactic dynamical system can be used for
testing the efficiency and reliability of other dynamical spectra
introduced in previous papers (see Zotos 2011a,b).  In particular, we
can check if these new definitions can identify resonant orbits of
higher multiplicity, different chaotic components and sticky regions
both in 2D and 3D systems.

The $S(k)$ spectrum is a very effective tool for the study of 3D motion,
as it produces as much spectra as is the number of islands of invariant
curves or, consequently, the number of of the 3D invariant manifolds.
Another advantage of the new spectrum is that it can be used for the
calculation of the 3D sticky motion period.  The $S(c)$ spectrum can be
also used to study the character of a 3D orbit, but it has no ability to
detect the 3D sticky motion.  The main reason is that the coupling of
the third component, $z$, carrying all the information about 3D motion
in general, is hidden in the definition of the $S(c)$ spectrum, but in
any case it affects the values of $x$, $p_x$ and $p_y$ entering relation
(5).  Using the definition of the $S(k)$ spectrum, we have managed to
overtake this minor drawback, and we have constructed a new dynamical
spectrum suitable for 3D orbits and especially for 3D sticky orbits.  We
must also point out, that when we deal with 2D motion, i.e. when
$(z=p_z=0)$, the $S(k)$ spectrum transforms to $S(c)$.

\thanks{I am much indebted to Professor N. D. Caranicolas for drawing my
attention to the study of 3D sticky orbits and for fruitful
discussions during this research.  I also thank Peeter
Tenjes for careful reading the manuscript and for important
suggestions and comments, which improved the quality of
the present paper.}

\vskip4mm

APPENDIX. {\bf The Taylor expansion}

\vskip4mm

In order to describe the motion in a triaxial elliptical galaxy, we
use the well known logarithmic potential
\begin{equation}
V_L=\frac{\upsilon _0^2}{2}\ln\left[x^2+ay^2+bz^2+c^2\right] \ ,
\end{equation}
where $\upsilon _0$ is used for the consistency of the galactic units,
$a$ and $b$ are flattening parameters and $c$ is the scale
parameter of the elliptical galaxy (see Caranicolas \& Zotos 2011).
Expanding potential $V_L$ in Taylor series about the origin and
keeping terms up to the second degree in the variables we find
\begin{eqnarray}
V_T &=&
\frac{\upsilon_0^2}{2}\ln\left[\left(\frac{x^2+ay^2+bz^2}{c^2}+1\right)c^2\right] =
\frac{\upsilon_0^2}{2}\ln\,c^2 +
\frac{\upsilon_0^2}{2c^2}\left(x^2+ay^2+bz^2\right) \nonumber \\
&=& \upsilon_0^2\ln\,c +
\frac{\upsilon_0^2}{2c^2}\left(x^2+ay^2+bz^2\right) \ .
\end{eqnarray}

Thus, the harmonic potential $V_T$ which is derived from the
logarithmic expansion is
\begin{equation}
V_T = \frac{\omega^2}{2}\left(x^2+ay^2+bz^2\right) \ ,
\end{equation}
where it was assumed that
\begin{equation}
\frac{x^2 + a y^2 + b z^2}{c^2} \ll 1 \ ,
\end{equation}
and
\begin{equation}
\omega = \frac{\upsilon_0}{2} \ .
\end{equation}

We use the values:  $\upsilon_0=20$, $a=4$, $b=1.25$, $c=2$.  Here we
must point out that these values of the involved parameters are valid
only for distances $R=\sqrt{x^2+y^2+z^2} \leq 1$ from the galactic
center.  We choose these particular values in order to describe and
study only the central parts of the galaxy, where the sticky regions and
the resonant phenomena take place.  From the experience of our previous
work we believe that typical values of the scale parameter, $c$, for an
elliptical galaxy are in the range $0.5 \leq c \leq 2.5$.

\References

\refb Barth A. J., Filippenko A. V., Moran E. C. 1999, ApJ, 525, 673

\refb Barth A. J., Ho L. C., Filippenko A. V. et al. 2001, ApJ, 546, 205

\refb Binney J., Tremaine S. 2008, {\it Galactic Dynamics},
Princeton Series in Astrophysics, 2nd ed.

\refb Caranicolas N. D., Papadopoulos N. I. 2007, AN, 328, 556

\refb Caranicolas N. D., Zotos E. E. 2010, New Astronomy, 15, 427

\refb Caranicolas N. D., Zotos E. E. 2011, Research in Astronomy and
Astrophysics, 11(7), 811

\refb Cincotta P. M., Giordano C. M., Perez M. J. 2006, A\&A, 455, 499

\refb Hasan H., Norman C. A. 1990, ApJ, 361, 69.

\refb Hasan H., Pfenniger D., Norman C. 1993, ApJ, 409, 91

\refb Ho L. C., Filippenko A. V., Sargent W.\,L.\,W. 1995, ApJS, 98, 477

\refb Ho L. C., Filippenko A. V., Sargent W.\,L.\,W. 1997, ApJS, 112,
315

\refb Ho L. C., Rudnick G., Rix H. W. et al. 2000, ApJ, 541, 120

\refb Kandrup H. E., Sideris I. V. 2002, CeMDA, 82, 61

\refb Kandrup H. E., Siopis Ch. 2003, MNRAS, 345, 727

\refb Kaneda H., Onaka T., Sakon I. 2005, ApJ, 632, L83

\refb Lauer T. R., Faber S. M., Gebhardt K. et al. 2005, AJ, 129, 2138

\refb Lichtenberg A. J., Lieberman M. A. 1992, {\it Regular and
Chaotic Dynamics}, Springer, 2nd ed.

\refb Maoz D., Nagar N. M., Falcke H., Wilson A. S. 2005, ApJ, 625, 699

\refb Nagar N. M., Falcke H., Wilson A. S. 2005, A\&A, 435, 521

\refb Saito N., Ichimura A. 1979, in {\it Stochastic Behavior in
Classical and Quantum Hamiltonian Systems}, eds.  G. Casati \& L. Ford,
Springer, p.\,137

\refb Shields J. C., Rix H. W., McIntosh D. H. et al. 2000, ApJ, 534,
L27

\refb Skokos Ch. 2001, J. Phys. A, 34, 10029

\refb Skokos Ch., Bountis T. C., Antonopoulos Ch. 2007, Physica D, 231,
30

\refb Zotos E. E. 2011a, New Astronomy, 16, 391

\refb Zotos E. E. 2011b, Baltic Astronomy, 20, 75

\end{document}